\documentclass[9pt]{extarticle}

\usepackage{url} 
\usepackage{soul}

\usepackage{empheq}
\usepackage[most]{tcolorbox}
\newtcbox{\mymath}[1][]{%
    nobeforeafter, math upper, tcbox raise base,
    enhanced, colframe=black,
    colback=black!8, boxrule=0.75pt,
    #1}
\usepackage{mathtools}
\usepackage{multirow}
\usepackage{adjustbox}

\usepackage{amsmath}
\usepackage{xfrac}
\usepackage{ragged2e}
\justifying
\newcommand{\Overbrace}[2]{{\overbrace{#2}^{#1}}}
\usepackage{multicol}
\usepackage{amssymb}

\NewDocumentCommand{\overarrow}{O{=} O{\uparrow} m}{%
  \overset{\makebox[0pt]{\begin{tabular}{@{}c@{}}#3\\[0pt]\ensuremath{#2}\end{tabular}}}{#1}
}
\NewDocumentCommand{\underarrow}{O{=} O{\downarrow} m}{%
  \underset{\makebox[0pt]{\begin{tabular}{@{}c@{}}\ensuremath{#2}\\[0pt]#3\end{tabular}}}{#1}
}

\usepackage[round]{natbib}

\usepackage{mathrsfs}
\usepackage{algorithm}
\usepackage{algpseudocode}

\usepackage{amsmath}
\usepackage[most]{tcolorbox}

\tcbset{colback=white, colframe=black, 
        toprule=0.7pt,
        leftrule=0.7pt,
        rightrule=0.7pt,
        bottomrule=0.7pt,
        bottom=10pt,
        top=2pt,
        left=10pt,
        right=10pt,
        sharp corners
        }

\def\notation{\bgroup
\section*{Notation}
\begin{description}\setlength{\itemsep}{-3pt} 
\def\notation##1{\item[##1]}}
\def\endnotation{\end{description}\vskip12pt\egroup}

\usepackage{mathpazo} 

\usepackage{geometry}
\usepackage{titlesec}
\titleformat{\section}[block]{\normalfont\large\bfseries}{\thesection}{1em}{}
\date{}

\usepackage{authblk}

\title{\textbf{\LARGE Hydrodynamics of fault gouges\\ from constitutive modelling to the physics of friction}}

\small{\author[1,2]{Filippo Masi}
\author[1]{Itai Einav}
\affil[1]{\itshape Sydney Centre in Geomechanics and Mining Materials, School of Civil Engineering, The University of Sydney, NSW 2006, Sydney, Australia.
\vspace{0.2cm}
}

\affil[2]{\itshape Univ. Grenoble Alpes, Inria, CNRS, Grenoble INP, Institute of Engineering, LJK, 38000 Grenoble, France.}
}
\begin{document}
\maketitle

\section*{Abstract}
\noindent 
The development of rate- and state-dependent friction laws offered important insights into the key physical mechanisms of the frictional behaviour of fault gouges and their seismic cycle. However, past approaches were specifically tailored to address the problem of fault shearing, leaving questions about their ability to comprehensively represent the gouge material under general loading conditions.
This work establishes an alternative approach for developing a physical friction law for fault gouges that is grounded on the rigour of the hydrodynamic procedure with two-scale temperatures through \emph{Terracotta}, a thoroughly robust constitutive model for clay in triaxial loading conditions. By specifying the model for direct shearing, the approach yields an alternative friction law that readily captures the frictional dynamics of fault gouges, including explicit dependencies on gouge layer thickness, normal stress, and solid fraction. Validated against available laboratory experiments, the friction law retains the original predictive capabilities of Terracotta in triaxial conditions and explains the rate-and-state, dilatational behaviour of fault gouges in direct shear conditions. Finally, when the Terracotta friction law is connected to a spring-dashpot representation of the host rock, the combined model predicts an elastic buildup precursor to the onset of and subsequent seismicity, with results closely reflecting experimental evidence and field observations.
While this study focuses on clay-rich gouges, the approach and findings are expected to offer much wider implications to a variety of materials.

\section*{Plain Language Summary}
\noindent Understanding how faults in the Earth's crust move and generate earthquakes is essential for assessing seismic hazards. Friction laws in geophysics describe the resistance that fault zones effectively transmit through gouge material. While existing friction laws have provided valuable insights, these are typically engineered directly to the specific conditions of sheared faults.
This study introduces a new friction law for sheared fault gouges, which is derived from first principles using an already existing description of clay materials. Originally designed to describe the complex behaviour of clay under conditions different than sheared fault gouges, the model is first generalised and is then specified to study the shear dynamics of earthquake fault zones. Thanks to the new approach, the model readily considers key factors such as gouge layer thickness, stress, and solid fraction. The new friction law accurately reproduces experimental and field observations, including critical phenomena such as fault build-up prior to a first earthquake and the post seismic cycle, all while revealing the evolution of dilatancy (volume changes).
Although this research focuses on clay-rich gouges, the approach can be extended to other materials, offering broader implications for understanding earthquake mechanics and fault zone processes.
\medskip

\section{Introduction}

Over recent decades, experimental and modelling efforts have focused on representing the frictional behaviour of faults to deepen our understanding of seismogenesis and the seismic cycle. Laboratory tests on rock, clay, and granular fault gouges have motivated the development of rate-and-state friction laws \citep{dieterich1979modeling, ruina1983slip, marone1998effect,saffer2003comparison}, which aim to describe key features of earthquake nucleation and occurrence. Originally developed to phenomenologically capture the frictional and healing behaviour observed in laboratory tests \citep{dieterich1979modeling}, these friction laws have been instrumental in linking experimental data with the mechanics of earthquakes and faulting \citep[e.g.,][]{scholz2019mechanics}. Rate-and-state frictional parameters are generally obtained through fitting the friction law to curves obtained from velocity stepping or slide-hold-slide friction experiments \citep{marone1998effect,saffer2003comparison}. The corresponding rate-and-state friction laws are conventionally adopted to model the mechanics of earthquakes and faulting, offering a robust explanation for numerous laboratory observations \citep{gu1984slip,gu1991}. Combined with continuum elasticity of the surrounding rock mass, these laws have also been applied to simulate natural and induced fault behaviours such as earthquake nucleation \citep{lapusta_rice_2000,scholz2019mechanics}, aftershock sequences \citep{segall_rice,95GL03076,chenlapusta2009}, postseismic or interseismic fault creep and healing \citep{marone1998effect,hetland2010}.

Nevertheless, several limitations remain. The main challenge in using rate-and-state friction laws is the empirical nature of their parameters, which do not directly correspond to specific physical mechanisms. Consequently, existing rate-and-state friction laws cannot readily explain fault responses under diversely different laboratory and natural conditions without constant adaptation of the material constants \citep[see][]{spiers2016}. In particular, earthquake nucleation and rupture propagation \citep{riceruina83, tserice1986} depend considerably on factors such as temperature, stress level and slip velocity, which complicates the calibration of rate-and-state friction laws \citep{spiers2016, bedford2022fault, carpenter2014}.
These limitations of rate-and-state friction laws stem from two main factors: (1) their empirical foundation, which is based on a rather retrospective and idealised view of contact asperity behaviour, and (2) the fact they ignore the gouge volume within the fault, which does not only sustain shear sliding but also consolidation and shear-induced dilation processes \citep[for further details, see][]{segall_rice,spiers2016}.

To address these limitations, recent effort has focused on the development of models inspired by physics at the microscale that capture steady-state and transient fault friction while grounding parameter constraints on micromechanics \citep[among others, see][]{niemeijer2007spiers,noda2008,daub_carlson,spiers2016,ikarietal2016,Niemeijer2017,perfettini2017,van2018comparison,barbot2019slow,barbot2022rate,pranger2022rate}. For instance, \citet{daub_carlson,daub2010pulse} developed a rate-and-state constitutive model rooted on the shear transformation zone theory \citep[see]{falk1998dynamics,bouchbinder2009nonequilibrium,lieou2012nonequilibrium} providing an effective description of dilation and compaction phenomena in the frictional behaviour of gouges. A pivotal development along this direction is the model by Chen, Niemeijer, and Spiers, also known as the CNS model, which incorporates deformation mechanisms at the microstructural level that differentiate between a localised shear band and a bulk gouge layer. With some adaptation, the model could further include normal stress effects on the frictional behaviour \citep{spiers_sigman}. To achieve these feats, unlike phenomenological rate-and-state friction laws, the model does not ignore the dilatational shear of faults \citep{einat2011friction}.

While micromechanically-inspired fault models such as the CNS model have made significant strides, their ability to represent the response of the same gouge material under loading conditions that differ from direct fault shear has not been tested, leaving questions about their ability to robustly represent the material. For example, a general physical model should be capable of representing the response of the same gouge material under isotropic loading, triaxial shear conditions, or even the dependence of fault friction on the reversal of the slip velocity that may happen during an earthquake. While specifically tailored for fault shearing, such cyclical conditions should engage gouge elasticity that is not accounted for in both rate-and-state and micromechanically-inspired friction laws. Similarly, most existing fault friction laws tend to focus exclusively on inelastic slip behaviour, ignoring the initial elastic response of the material prior to the establishment of mature shear response. Therefore, the scope of these previous models cannot be easily extended to explain questions regarding the genesis of frictional faults. 
Finally, it is also important to note that these models have not fully addressed how the frictional behaviour depends on the thickness of the slip layer, despite experimental and in-situ evidence for this effect \citep[e.g.,][among others]{byerlee1976note, 93JB03361, 2011JB008264, evans1990thickness, JB095iB05p07007, einat2011friction,scuderi2014physicochemical, lyu2019mechanics, bedford2021role}.

\begin{figure}[h]
\centering
\includegraphics[width=\textwidth]{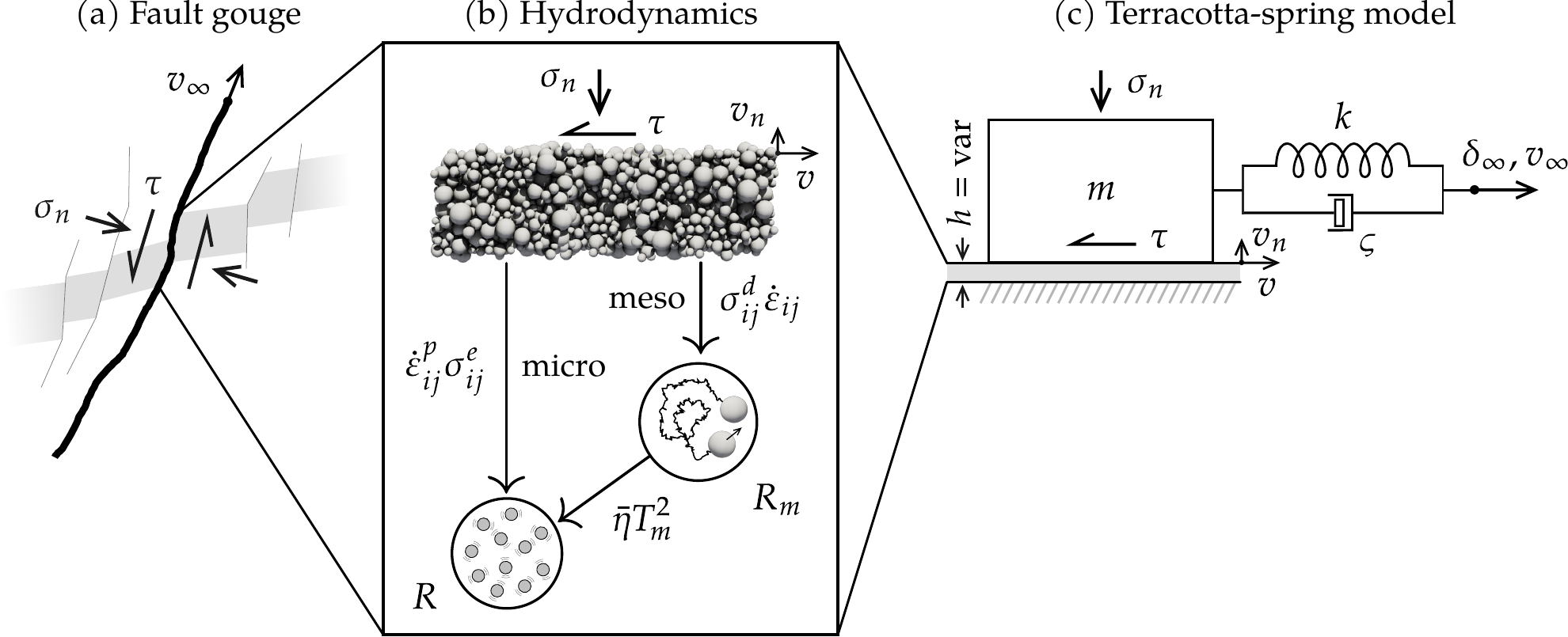}
\caption[]{Hydrodynamics of fault gouges: from constitutive to friction law. A fault gouge (a) subjected to tectonic movement with slip velocity $v_{\infty}$ is represented as (b) a slip layer of thickness $h$, subjected to an overburden stress $\sigma_n$, experiencing slip and normal velocities (known as rates), $v$ and $v_n$, that control the development of the shear stress $\tau$. The internal energy of the gouge is either held in the system through elastic stress, $\smash{\sigma^e_{ij}}$, or dissipated as a two-stage irreversible mechanism through the generation of thermal and meso-related entropy productions  ($R$ and $R_m$) as functions of the plastic strain rate, $\smash{\dot{\varepsilon}^p_{ij}}$, and the viscous stress, $\smash{\sigma^d_{ij}}$, both dependent on the meso-temperature, $T^m$, while $\bar{\eta}$ is a hydrodynamic coefficient that controls the energy decay between the scales. This hydrodynamic model is attached to a spring-dashpot system (c), with a spring of stiffness $k$ and a dashpot of damping ratio $\varsigma$ representing the host rock, to predict the dynamics of earthquakes, in clay-rich fault gouges. }
\label{fig:spring-slider}
\end{figure}

The objective of this work is the development of a new physics-based friction law for fault gouges that is grounded upon hydrodynamic principles. The key idea of this work lies in the philosophy of the development (see Figure \ref{fig:spring-slider}). Instead of constructing a law directly by inspecting the specific loading conditions the gouge material sustains while the fault slides, we use an existing constitutive model for clay \citep{wiebicke2024simple}, originally developed for triaxial loading conditions that are different to those during fault sliding. This approach ensures that our results are objective and free from potential bias. The model named \emph{Terracotta}, leverages the rigorous hydrodynamic procedure by \citet{landau2013statistical}, which was developed originally for fluid continua. Later, this framework was extended to describe liquid crystals \citep{de1993physics}, granular media \citep{jiang2009granular}, and partially saturated porous media \citep{einav2023hydrodynamics}, amongst other materials. Having been grounded upon hydrodynamic principles, Terracotta satisfies the first and second law of thermodynamics, momentum and mass balance laws \citep{einav2018hydrodynamic}, material objectivity, and the thermodynamic reciprocity of irreversible thermodynamic fluxes \citep{onsager1931reciprocal}, as well as considers the physical scale separation of temperatures in heterogeneous media and the energy flow in between \citep{jiang2009granular}.

Building on the original Terracotta model for triaxial loading conditions, we first construct the tensorial generalisation of the model, while preserving all of its original physics and predictive triaxial features. Equipped with the tensorially generalised model, we move further by specifying the unique boundary conditions of an isolated fault gouge under direct shear loading. In so doing, we derive an alternative friction law that automatically preserves the model's capacity to represent volumetric material response, including during consolidation and dilatational shear. The resulting physics-based friction law encapsulates the essential features of both rate-and-state and micromechanically-inspired friction laws while naturally introducing dependencies on key meaningful parameters and state variables.

The resulting \emph{Terracotta friction law} incorporates several physical phenomena previously ignored by fault friction laws. For example, the model accurately represents the pressure- and density-dependent elasticity of gouge media, allowing it to simulate the buildup of an elastic stress, which fully lets it account for the effects of the total normal stress. The Terracotta friction law also builds on the concept of critical state in soil mechanics \citep{wood1990soil} and the thermodynamic reciprocity of plastic volumetric and deviatoric plastic straining \citep{wiebicke2024simple}. The latter feature enhances the capability of the model to predict post-peak frictional behaviour, providing a physically grounded explanation for dilatational slip hardening and weakening.
An important feature of the friction law is the consideration of a finite fault gouge thickness as a physical quantity. Its temporal evolution directly depends on the prescribed normal stress under direct shear boundary conditions, which also influences the elastic development of frictional instabilities. Consistent with the original Terracotta model \citep{wiebicke2024simple}, the derived friction law accounts for the influence and evolution of the porosity (or alternatively, the solid fraction) of the gouge material.
Another key ingredient is the inclusion of the meso-related temperature (from here, to be called \emph{meso-temperature}) as a non-equilibrium variable for clay, rooted on prior developments for granular media \citep{jiang2009granular,alaei2021hydrodynamic}. Unlike the conventional thermal temperature, which reflects atomic-scale thermal fluctuations, the meso-temperature captures the kinetic energy of the velocity fluctuations of the clay mesostructures, including the motion of platelets and aggregates that are overwhelmingly larger than atoms, thus attributed to have their own temperature. When stressed or deformed, these mesostructures exhibit fluctuations akin to granular particles, with pores expanding, collapsing, and clusters fragmenting. This motion at the mesoscale influences the material's elasticity and viscosity, as depicted in Figure \ref{fig:spring-slider}. Despite the advanced hydrodynamic procedure, the Terracotta friction law reveals a simple mathematical structure (as shall eventually be captured mathematically in Equation (\ref{eq:tau_specified})).

Accounting for these state variables and physical mechanisms enables the study of earthquake nucleation and occurrence by integrating the Terracotta friction law with conventional spring-dashpot model of the host rock, see Figure \ref{fig:spring-slider}. The resulting Terracotta-spring model naturally incorporates the evolution and dependency of the slip zone thickness and normal stress. This model is then used to simulate the full seismogenesis cycle, including the buildup of shear stress and deformations, followed by the creation of frictional instabilities and their evolution towards limit cycles. Through a detailed parametric analysis of the key physical parameters of the Terracotta friction law, we study and quantify the influence of the: (1) slip zone thickness, (2) solid fraction, (3) normal stress, and (4) far-field velocity. The results demonstrate the predictive capabilities of this friction law in accurately capturing dependencies of the frictional behaviour, in agreement with natural and induced seismicity data, previous experimental studies on clay-rich fault gouges, and discrete element method numerical simulations.

The paper is organised in a way that aims to progressively unfold the physical origin and rationale underlying each quantity and parameter in Figure \ref{fig:spring-slider}, followed by a rigorous assessment of their validity. Section \ref{sec:model} introduces the physical principles guiding the hydrodynamic procedure, before briefly presenting the Terracotta constitutive model, originally formulated for triaxial testing conditions. The Terracotta model is then generalised tensorially to enable full description of any conceivable loading conditions, before being reduced to the specific loading conditions of direct shearing of fault gouges.
In Section \ref{sec:validation}, we demonstrate the predictive capabilities of the resulting Terracotta friction law in predicting frictional stress-strain behaviours, including the elastic buildup of the shear stress, response under velocity stepping, and performance under different normal stresses, all as have been observed in laboratory experiments of clay-rich fault gouges.
After the validation of the Terracotta friction law against controlled velocity experiments, Section \ref{sec:spring-slider} integrates the Terracotta friction law with a spring-dashpot analogue of the host rock to investigate fault genesis and quantify the impact of the key physical parameters on earthquake dynamics. In conclusion, we discuss the broader implications of the new model, considering its hydrodynamic foundations and potential to explore non-local features for the spatial and temporal modelling of faulting.

\section{The model}
\label{sec:model}
\begin{figure}[ht]
\centering
\includegraphics[width=\textwidth]{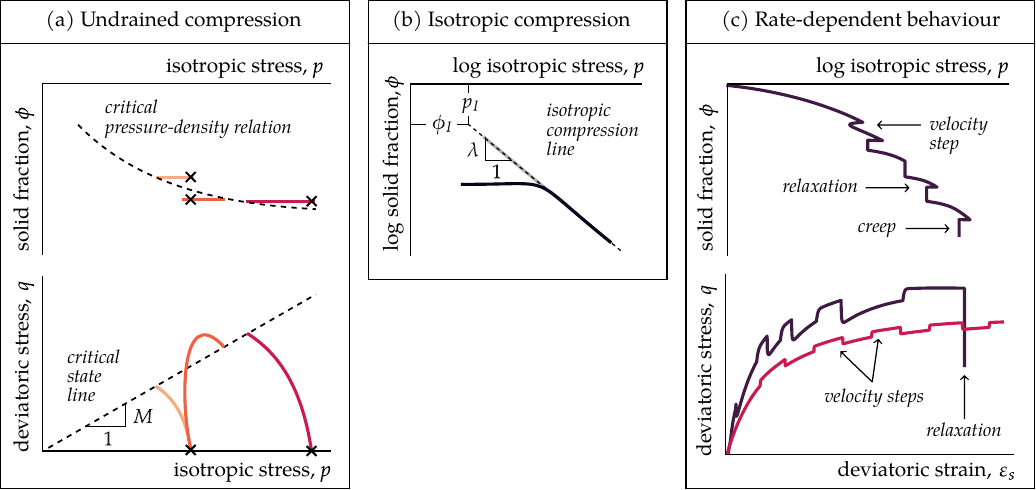}
\caption{Schematic outline of the range of behaviours captured by the Terracotta constitutive model in its original triaxial form (re-adapted from Wiebicke \& Einav, 2024). The panels showcase the model's predictions for: (a) undrained triaxial compression, (b) isotropic compression, and (c) rate-dependent processes in uniaxial compression (top) and undrained triaxial compression (bottom).}
\label{fig:scheme_triaxial}
\end{figure}

In striving to capture the rich, rate-dependent mechanical processes in clays, an abundance of constitutive models has been developed in geomechanics, mostly founded purely on curve fitting experimental data through a variety of empirical mathematical formulations. Recently, an alternative physics-based constitutive model, called \emph{Terracotta} \citep[see][]{wiebicke2024simple}, was proposed through the rigorous hydrodynamic procedure equipped with two physical temperatures separating the scale of atoms from that of the meso-structures of clay particles and aggregates. Unlike earlier constitutive models, Terracotta captures major features of the mechanical behaviour of clays under triaxial loading conditions, while avoiding to rely on abstract concepts from the mathematical theory of plasticity \citep[cf.][]{alaei2021hydrodynamic,riley2023constitutive}. The capacity of the model to describe an astounding range of phenomena is schematically represented in Figure \ref{fig:scheme_triaxial}, showcasing its application to both undrained triaxial and isotropic compression, which is consistent with empirical observations of critical state in soil mechanics (see insets a and b). Additionally, Terracotta accurately predicts an impressive range of rate-dependent experimental observations, such as velocity stepping effects, relaxation, and creep phenomena (cf. inset c).

Relative to past inelastic constitutive models for clay, and despite being strongly rooted in physics, Terracotta possesses a highly simple mathematical structure that motivates studies in geophysics. Here, we therefore seek to apply this model to direct fault shearing conditions. To this end, we present hereinafter the theoretical background behind the core of the hydrodynamic procedure and outline the Terracotta model, in its original form for triaxial testing conditions. We then tensorially generalise the model to accommodate any boundary and loading conditions. We then apply the generalised model to the specific case of direct fault shearing. The resulting equations establish a new friction law for clay-rich fault gouges, which we call \emph{Terracotta friction law} in the following, as conceptualised in Figure \ref{fig:friction}. 

\subsection{Background setting}
\label{subsec:back}
\subsubsection{Hydrodynamics and two-stage irreversibility}
\label{subsec:hydro}

In the following we summarise the main hydrodynamic principles articulated by \citet{landau2013statistical,jiang2009granular, einav2023hydrodynamics}, on which Terracotta was founded upon. As such, unlike past constitutive models for clay, the development of Terracotta \citep{wiebicke2024simple} has benefitted from :
\begin{enumerate}
    \item the conservation of energy, momentum and mass; 
    \item non-negativity of the entropy production;
    \item scale separation of temperatures in heterogeneous media;
    \item understanding of energy flow across the corresponding scales;
    \item thermodynamic reciprocity \citep{onsager1931reciprocal} of volumetric and deviatoric plastic fluxes;
    \item stationarity of irreversible fluxes close to thermodynamic equilibria; and
    \item the relationship between thermodynamic pressure and density.
\end{enumerate}

A key feature of the model is the incorporation of the meso-temperature as a non-equilibrium state variable in clay. Unlike the conventional thermal temperature, which represents the kinetic energy of the fluctuating motions of (microscopic) atoms, the meso-temperature accounts for the kinetic energy of the fluctuating motions of (mesoscopic) clay aggregates and platelets that are overwhelmingly larger than the atoms. Through the two-stage irreversibility principle of \citet{jiang2009granular}, the two temperatures are used to describe the energy flow within the material from the continuum scale down to the atomistic microscale, through the mesoscale of clay aggregates and platelets.

The concept of meso-temperature, and its thermodynamic conjugated meso-entropy, is designed to capture a broad spectrum of mesoscopic states, including non-affine rearrangements and dynamic responses \citep{goldhirsch2008introduction,jiang2017granular,liu2021temperatures,maranic2021granular,alaei2021hydrodynamic}. This is in contrast with the concept of effective temperature \citep{falk1998dynamics,bouchbinder2009nonequilibrium} -- derived from Edward's configurational entropy and compactivity \citep{edwards1989theory} -- which addresses only a limited subset of configurational states, typically those associated with near-equilibrium or quasi-static conditions. Accounting for the meso-temperature is particularly important for systems under transient loadings, such as those occurring within fault gouges during transitions from stick to slip motions and vice versa. 

Although not as directly measurable as its thermal counterpart, the meso-temperature can be inferred by (1) analysing the fluctuating trajectories of meso-structures \citep[e.g., particles in sand, and platelets or aggregates in clay, cf.][]{maranic2021granular}, with advanced microscopic techniques or high-resolution simulations, or by (2) relating it to kinetic pressure \citep[e.g.][]{goldhirsch2008introduction,alaei2021hydrodynamic}. 

Following \citet{wiebicke2024simple}, the state of clay materials is characterised by bulk density, $\rho$, elastic strain, $\varepsilon^e_{ij}$, and thermal and meso-related entropies, $s$ and $s_m$, which are the thermodynamic conjugates of the thermal and meso-temperatures, respectively. Accordingly, the general form of the internal energy density is defined by
 
\begin{equation}
    u \equiv u \left( \rho, \varepsilon^e_{ij}, s, s_m \right).
    \label{eq:general_u}
\end{equation}
 
The complete derivation of the hydrodynamic procedure can be found in \citet{alaei2021hydrodynamic} for materials that involve exactly the same set of state variables. Following the same procedure, the total stress is defined as
 
\begin{equation}
    \sigma_{ij} \equiv \sigma_{ij}^e+\sigma^d_{ij} +p^T\delta_{ij}, \qquad \sigma^e_{ij} \equiv \frac{\partial u}{\partial \varepsilon_{ij}^e}, \quad p^T \equiv - \frac{\partial \left( u/\rho \right)}{\partial \left( 1/\rho \right)}\Bigg|_{\varepsilon_{ij}^e, \frac{s}{\rho}, \frac{s_m}{\rho}},
    \label{eq:stress_hydrodynamics}
\end{equation}
 
where $\sigma^e_{ij}$ is the elastic stress tensor, $p^T$ is the thermodynamic pressure, and $\sigma^d_{ij}$ is the dissipative viscous stress, with $\delta_{ij}$ being the Kronecker delta.

Next, we determine the evolution equations for the four state variables embodied in Equation (\ref{eq:general_u}). Noting that throughout the article, Einstein’s summation is implied for repeated indices $i,j=1,2,3$, the evolution of the bulk density is derived from the mass balance, $\dot{\rho}+\nabla_i\left(\rho v_i\right) = 0$, with $v_i$ being the velocity vector. We adopt the usual sign convention in soil mechanics that consider positive deformation in compression and identify the trace of the velocity gradient with the negative volumetric strain rate, $\nabla_i v_i = - \dot{\varepsilon}_{kk} \equiv -\dot{\varepsilon}_v$. By further considering dry fault gouges with negligible air mass relative to the solid mass, the solid fraction could be calculated using $\phi \equiv \rho/\rho_s$, where $\rho_s$ is the solid density defined as the ratio of the solid mass over the solid volume. The evolution equation for the solid fraction is then defined by
 
\begin{equation}
    \dot{\phi} + v_i \nabla_i \phi = \phi \dot{\varepsilon}_v,
    \label{eq:dot_phi}
\end{equation}
 
under the assumption of incompressible solid particles, i.e., $\dot{\rho}_s\approx 0$. Modifications that account for air and water densities within the hydrodynamic procedure would go beyond the current scope of this paper, but are certainly possible and would follow something akin to the work in \citep{chen2023hydrodynamic,chen2024hydrodynamic}.   

The evolution for the elastic strain is given by \citet{einav2018hydrodynamic} and reads
 
\begin{equation}
    \dot{\varepsilon}_{ij}^e+ v_k\nabla_k \varepsilon^e_{ij} +\Omega_{ik}\varepsilon^e_{kj} -\varepsilon^e_{ik}\Omega_{kj}= \dot{\varepsilon}_{ij} - \dot{\varepsilon}_{ij}^p, 
    \label{eq:dot_eps_el}
\end{equation}
 
where $\dot{\varepsilon}_{ij}$ and $\Omega_{ij}$ are respectively the symmetric and skew-symmetric parts of the spatial velocity gradient, $\nabla_j v_i$, and $\dot{\varepsilon}_{ij}^p$ is the inelastic strain rate. Following scale separation \citep{jiang2009granular}, where the granular entropy is replaced by the more general meso-related one~\citep{einav2023hydrodynamics,wiebicke2024simple}, the thermal and meso-related entropies satisfy the following balance equations:
 
\begin{subequations}
\begin{align}
    \label{eq:dot_s}
    \dot{s} +\nabla_i\left(sv_i-f_i \right)&= \frac{R}{T}\geq 0, \\
    \dot{s}_m +\nabla_i\left(s_mv_i-f^m_i \right)&= \frac{R_m}{T_m},
    \label{eq:dot_sm}
\end{align}
\label{eq:dot_s_sm}
\end{subequations}
 
where $T\equiv \frac{\partial u}{\partial s}$ and $T_m \equiv \frac{\partial u}{\partial s_m}$ are the thermal and the meso-temperatures, respectively; $sv_i$ and $s_mv_i$ are the thermal and meso-related convective entropy currents; $f_i$ and $f^m_i$ are the thermal and meso-related dissipative entropy currents; and $R$ and $R_m$ are the thermal and meso-related entropy productions. Note that only the former is required to be positive-semidefinite by virtue of the second law of thermodynamics.

By combining Equations (\ref{eq:dot_phi}-\ref{eq:dot_s_sm}) with the momentum and energy balances, and the principle of two-stage irreversibility in the seminal work by \citet{jiang2009granular}, the total rate of entropy production comprises two-scale contributions:
 
\begin{subequations}
\begin{align}
R+R_m &\geq 0\\
R_m & = \sigma^d_{ij}\dot{\varepsilon}_{ij} +f^m_i\nabla_i T_m - \bar{\eta}T_m^2,\\
R & = \dot{\varepsilon}_{ij}^p \sigma_{ij}^e +f_i\nabla_i T +\bar{\eta}T_m^2\geq 0,
\end{align}
\label{eq:entropy_production}
\end{subequations}
 
where the viscous dissipation $\sigma^d_{ij}\dot{\varepsilon}_{ij}$ is entirely attributed to the meso-related entropy production, unlike \citet{jiang2009granular} who placed part of it in the thermal entropy.
The dissipated energy within the system can either agitate the microstructure, thereby elevating the thermal temperature, $T$, or the mesostructures, thus elevating $T_m$, while overtime the energy can flow from the meso- to the micro-scale in a two-stage process, see Figure \ref{fig:spring-slider}(b). The flow between the scales is given by the $\pm \bar{\eta}T_m^2$ sink/source terms, with $\bar{\eta}\geq 0$ so that the energy will indeed flow from the higher to the lower scale. 

Next, we look at the dissipative fluxes in the entropy production (\ref{eq:entropy_production}), namely: $f_i$, $f_i^m$, $\sigma^d_{ij}$, and $\dot{\varepsilon}_{ij}^p$ and determine their hydrodynamic equations in terms of Onsager's reciprocity relationships \citep{onsager1931reciprocal}. 
Following \citep{wiebicke2024simple} these are specified by
 
\begin{eqnarray}
    \label{eq:viscous_stress}
    f_i = \frac{k_T}{T}\nabla_i T, \qquad f_i^m = \frac{k_m}{T_m}\nabla_i T_m,\qquad
    \sigma^d_{ij} = \mathbb{X}^d_{ijkl} \dot{\varepsilon}_{kl}, \qquad 
    \dot{\varepsilon}^p_{ij} = \mathbb{X}^p_{ijkl} \sigma^e_{kl},
\end{eqnarray}
 
where $k_T$ and $k_m$ are the micro- and meso-scopic diffusivity coefficients, $\mathbb{X}^p_{ijkl}$ and $\mathbb{X}^d_{ijkl}$ are the transport coefficients, which could in general be functions of the state variables.
Following \citet{onsager1931reciprocal}, the transport coefficients must fulfil a number of reciprocity requirements stemming from time-reversal symmetry and the entropy production inequality (\ref{eq:entropy_production}), as described by \citet{wiebicke2024simple}. 

\subsubsection{Original triaxial form of the model}
\label{subsec:triaxial}

Building upon the above theoretical and hydrodynamic settings, the Terracotta model \citep{wiebicke2024simple} was originally developed for standard triaxial conditions, in terms of triaxial stress and strain invariants. 
The isotropic stress and volumetric elastic strain invariants are defined as $p=\frac{\sigma_{ii}}{3}$ and $\varepsilon^e_v = \varepsilon^e_{ii}$, respectively. The deviatoric stress and elastic strain invariants read $q=(\sfrac{3}{2} s_{ij} s_{ij})^{\sfrac{1}{2}}$ and $\varepsilon^e_s = (\sfrac{2}{3} e^e_{ij} e^e_{ij})^{\sfrac{1}{2}}$, where $s_{ij} = \sigma_{ij}-p\delta_{ij}$ and $e^e_{ij} = \varepsilon_{ij}^e-\frac{\varepsilon_v^e}{3}\delta_{ij}$ are the deviatoric stress and elastic strain tensors, respectively.

The original form of the model considers isothermal processes, hence the dependency of the internal energy with respect to the thermal entropy is neglected. The internal energy density, $u$, in Equation (\ref{eq:general_u}), is decomposed into elastic and meso-related contributions, $u^e$ and $u^m$, through:
 
\begin{equation}
\label{eq:triaxial_u}
    u = u^e ( \phi, \varepsilon^e_{ij} )+u^m \left( s^m \right),\qquad 
    u^e =\phi^6\left(\frac{\tilde{K}}{6}{\varepsilon^e_{v}}^3+\frac{3}{2}\tilde{G}\varepsilon^e_{v} {\varepsilon_s^e}^2 \right),\qquad
    u^m =\frac{\Gamma}{4} {s_m}^2,
\end{equation}
 
where $\tilde{K}$ and $\tilde{G}$ represent the intrinsic bulk and shear stiffness constants, respectively; and $\Gamma= 1\text{ K\textsuperscript{2}/kPa}$ is a positive constant controlling the contribution of meso-related entropy to the internal energy (its inverse could be considered as a meso-related analogue of the thermal heat capacity).

\begin{figure}[ht]
\centering
\includegraphics[width=\textwidth]{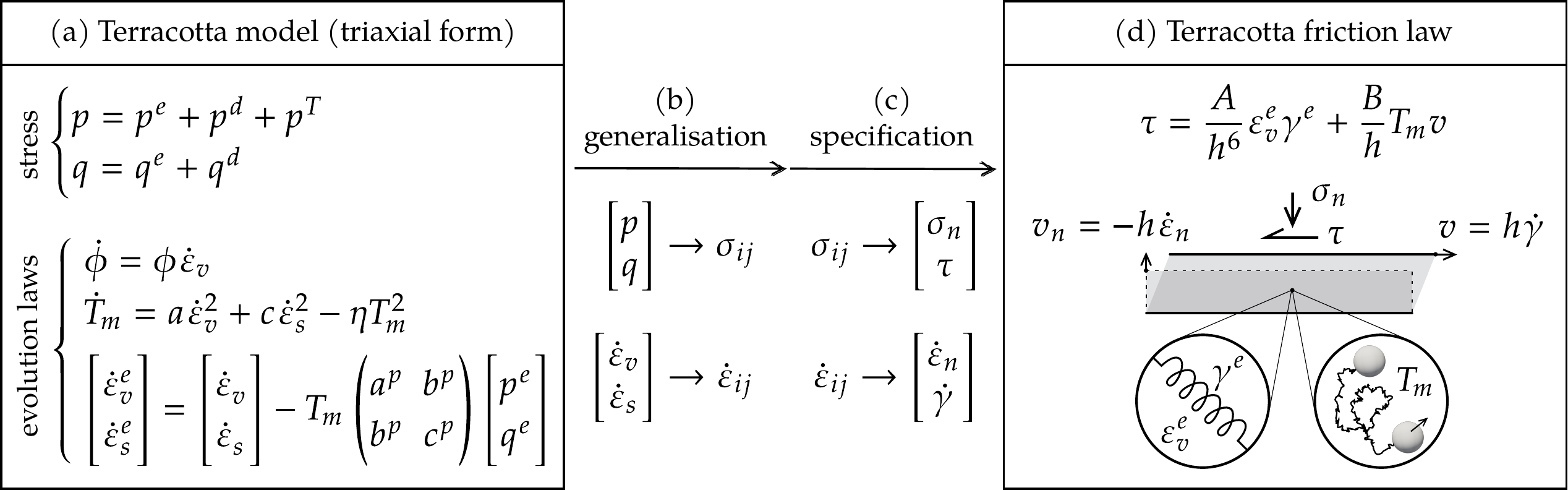}
\caption{Summary of the mathematical development from (a) the original Terracotta model for triaxial conditions to its (b) tensorial generalisation to its (c) specification for direct fault shearing conditions, which finally reduces to (d) the resulted Terracotta friction law. The total stress in Terracotta is equal to the sum of elastic, thermodynamic, and viscous contributions. The elastic stress (in invariant form $p^e$ and $q^e$), is expressed by Equation (\ref{eq:elastic_stress_invariants}); the thermodynamic pressure ($\smash{p^T}$) is defined according to (\ref{eq:stress_hydrodynamics}); while the viscous stress (in invariant form $\smash{p^d}$ and $\smash{q^d}$) is expressed by Equation (\ref{eq:pdqd}); and finally, the  plastic transport coefficients are given by Equation (\ref{eq:transport_coefficients_ap_T}). The resulting friction law for considers the slip velocity $v$, gouge thickness $h$, meso-temperature $T_m$, and elastic strains $\gamma^e$ and $\varepsilon_v^e$.}
\label{fig:friction}
\end{figure}

By virtue of the stress decomposition in Equation (\ref{eq:stress_hydrodynamics}), the volumetric and deviatoric stress invariants are given by
 
\begin{subequations}
\begin{align}
\label{eq:triaxial_stress__}
p &= p^e+p^d+p^T,\\
q &= q^e+q^d,
\end{align}
\end{subequations}
 
where $p^e, q^e$ and $p^d,q^d$ are respectively the elastic and dissipative (viscous) components. The elastic volumetric and deviatoric stresses and the thermodynamic pressure are obtained from Equation (\ref{eq:stress_hydrodynamics}), giving
 
\begin{subequations}
\label{eq:triaxial_stress}
\begin{align}
\label{eq:elastic_stress_invariants}
p^e &\equiv \frac{\partial u^e}{\partial \varepsilon^e_v}= \phi^6\left(\frac{\tilde{K}}{2}{\varepsilon_v^e}^2+\frac{3\tilde{G}}{2}{\varepsilon_s^e}^2\right),\\
\label{eq:elastic_stress_invariants_q}
q^e &\equiv \frac{\partial u^e}{\partial \varepsilon^e_s}= 3\tilde{G} \phi^6 \varepsilon_v^e \varepsilon_s^e,\\
\label{eq:elastic_stress_invariants_pT}
p^T & = \frac{T_m^2}{\Gamma} \Leftarrow T_m = \frac{\Gamma}{2}s^m,
\end{align}
\end{subequations}
 
where the elastic contribution in the thermodynamic pressure, i.e., $5u^e$, was found negligible  \citep{wiebicke2024simple} and is thus omitted for the sake of simplicity. The above equations are completed by the constitutive relationships for the dissipative fluxes, $p^d$ and $q^d$, and the evolution equations for the solid fraction, Equation (\ref{eq:dot_phi}), the meso-temperature, $T_m$, and the elastic strain, $\varepsilon_v^e$ and $\varepsilon_s^e$. Their derivation follows the hydrodynamic procedure, cf. subsection \ref{subsec:hydro}, and the exact procedure is sketched below for the generalised model in tensorial form. For the original derivation in triaxial conditions, we refer to \citet{wiebicke2024simple}, with the corresponding expressions presented in Figure \ref{fig:friction}(a). 

More precisely, the calibration of the triaxial invariants of the viscous stress, $p^d$ and $q^d$, is performed in the simplest way to ensure the semi-positivity of the viscous dissipation, leading to the following formulation:
 
\begin{equation}
p^d = aT_m\dot{\varepsilon}_v, \quad q^d = cT_m\dot{\varepsilon}_s,\label{eq:pdqd}
\end{equation}
 
where the viscous stress is made to decay to zero as the meso-temperature approaches thermodynamic equilibrium ($T_m\rightarrow 0$), as expected from non-equilibrium thermodynamic properties.

Next, the stationary value of the meso-temperature is used to calibrate the dependence of the plastic transport coefficients, so to enable the recovery of experimental critical state observations. This calibration approach was proposed by \citet{wiebicke2024simple} and was informed by two canonical asymptotic states: normal consolidation state emerging by continuous isotropic compression and the critical state emerging by continuous shearing \citep{wiebicke2024simple}. Following this procedure, the following plastic transport coefficients were obtained:
 
\begin{equation}
    a^p = \sqrt{\frac{\eta}{a}}\frac{1}{p_c(\phi)} \quad 
    b^p  = -a\frac{q^e}{M^2p^e},\quad 
    c^p = \sqrt{\frac{\eta}{c}}\frac{1}{M\omega p_c(\phi)}+\frac{a}{M^2}, \quad p_c(\phi) = p_I\phi^{\lambda},
\label{eq:transport_coefficients_ap_T}
\end{equation}
 
where $a$, $c$ are positive material constants, $\omega$ and $M$ are the parameters describing the critical state line \citep[see Figure \ref{fig:scheme_triaxial} and refer to][]{wiebicke2024simple}, and $ p_c(\phi)$ describes the stress-density relationship on the isotropic compression line, where $\lambda$ is the slope of the isotropic compression line on a bi-logarithmic plot of pressure against solid fraction and $p_I = \phi_I^{-\lambda}$ (kPa) is calibrated with $\phi_I$ representing the solid fraction at a reference pressure of $1$ kPa, see Figure \ref{fig:scheme_triaxial}.

\subsection{Generalisation to the tensorial form}
\label{subsec:tensorial}
Based on the hydrodynamic procedure, we extend the above form of the Terracotta model to its tensorial form with the objective of addressing more general loading conditions than just triaxial loading, including those related to the shearing of fault gouges. By following the specification of the internal energy density given in Equation (\ref{eq:triaxial_u}), we generalise the elastic contribution in terms of the full elastic strain tensor, $\varepsilon_{ij}^e$, namely
 
\begin{equation}
    u^e =\phi^6\left(\frac{\tilde{K}}{6}{\varepsilon^e_{v}}^3+\tilde{G}\varepsilon^e_{v} e^{e}_{ij} e^{e}_{ij} \right),
    \label{eq:energy_elastic_gen}
\end{equation}
 
which is equivalent to the original expression in Equation (\ref{eq:triaxial_u}) when specified in terms of triaxial invariants. As in the triaxial form of Terracotta, we neglect the dependence of the internal energy on the thermal temperature. However, such dependencies could easily be accounted for through the hydrodynamic procedure, as further discussed in paragraph \ref{subsubsec:heat_equation}.

By virtue of Equation (\ref{eq:stress_hydrodynamics}), the elastic stress tensor becomes
 
\begin{equation}
\sigma^e_{ij}  \equiv \frac{\partial u}{\partial \varepsilon_{ij}^e}= \phi^6 \Bigg( \frac{\tilde{K}}{2} \varepsilon^{e^2}_{v}\delta_{ij} + \tilde{G} \delta_{ij} e^{e}_{kl} e^{e}_{kl} +2\tilde{G} \varepsilon^e_{v} e^{e}_{ij}\Bigg),
\label{eq:elastic_stress}
\end{equation}
 
whose volumetric and deviatoric invariants are exactly the same as those in the triaxial formulation, i.e., Equations (\ref{eq:triaxial_stress}). The mesoscopic component of the internal energy, $u^m$, is kept the same as in Equation (\ref{eq:triaxial_u}), ensuring that the resulting thermodynamic pressure in Equation (\ref{eq:elastic_stress_invariants_pT}), $p^T$, also remains unchanged.

It is worth noticing that the elastic energy density in Equation \ref{eq:energy_elastic_gen} reveals a physical elastic instability when the determinant of the instantaneous stiffness matrix, i.e., the Hessian $\partial \sigma^e_{ij}/\partial \varepsilon^e_{kl}$, vanishes. The instability criterion is equivalent with that obtained in the original triaxial form \citep[see][]{wiebicke2024simple} and can be expressed in terms of the ratio between the deviatoric and volumetric elastic stresses:
 
\begin{equation}
    \frac{q_e}{p_e}\leq M_e, \qquad M_e = \sqrt{\frac{3\tilde{G}}{\tilde{K}}}.
    \label{eq:Me}
\end{equation}

\subsubsection{Thermal and meso-related heat equations}
\label{subsubsec:heat_equation}
The concept of two-stage irreversibility \citep{jiang2009granular} provides us with the form of the evolution equations for the micro- and meso-temperatures. The full hydrodynamic derivation of their corresponding rate equations can be found in \citep[Appendix A][]{riley2023constitutive}, considering the two-stage irreversibility. In the absence of convective entropy currents, the evolution equations for the micro- and meso-temperatures are given by
 
\begin{subequations}
\begin{align}
        \dot{T} &= \frac{k_T}{c_T\rho}\nabla^2_i T+\dot{\varepsilon}_{ij}^p\sigma^e_{ij}+\bar{\eta}T_m^2\\
        \frac{2T_m}{\Gamma}\dot{T}_m & = k_m\nabla^2_i T_m+\sigma^d_{ij}\dot{\varepsilon}_{ij}-\bar{\eta} T_m^2,
        \label{eq:dot_Tm_nonlocal}
\end{align}
\label{eq:dot_T_nonlocal}
\end{subequations}
 
where $c_T$ is the micro-related specific thermal heat capacity.
Note that the Laplace operator, $\nabla^2_i \cdot$, represents the diffusion of the thermal and meso-temperatures and can physically capture second-gradient phenomena, known as non-local effects in granular physics  \citep{kamrin2012nonlocal,henann2013predictive}. Predicting these phenomena requires spatial integration of the underlying boundary value problem, which is beyond the focus of the current paper. From here on, we shall therefore neglect the second gradient terms and focus on phenomena that happen homogeneously within representative elementary volumes, independent of non-local effects.

\subsubsection{Dissipative fluxes and calibration}
By virtue of the time-reversal symmetry requirements, the relationships for the viscous stress and plastic strain rate introduced in Equation (\ref{eq:viscous_stress}) are generalised in term of the volumetric invariant and the deviatoric tensor as follows,
 
\begin{equation}
\renewcommand\arraystretch{1.3}
    \begin{bmatrix}
    p^d\\
    s^{d}_{ij}
\end{bmatrix}
= \frac{2T^m}{\Gamma}
\begin{pmatrix}
    a^d & b^d_{kl}\\
    b^d_{ij} & c^d_{ijkl} 
\end{pmatrix}
\begin{bmatrix}
    \dot{\varepsilon}_v\\
    \dot{e}_{kl}
\end{bmatrix},\qquad 
\begin{bmatrix}
    \dot{\varepsilon}_v^p\\
    \dot{e}^p_{ij}
\end{bmatrix}
= T_m
\begin{pmatrix}
    a^p & b^p_{kl}\\
    b^p_{ij} & c^p_{ijkl} 
\end{pmatrix}
\begin{bmatrix}
    p^e\\
    s^e_{kl}
\end{bmatrix}.
\label{eq:plastic_rate_coeff}
\end{equation}

Next, following \citet{wiebicke2024simple}, we consider the energy sink coefficient to be linearly proportional to the meso-temperature, i.e., $\bar{\eta} = 2\frac{\eta}{\Gamma}T_m$. We then substitute the constitutive relationship for the viscous stress in the meso-related heat equation (\ref{eq:dot_Tm_nonlocal}), so by neglecting non-local effects we obtain the following evolution equation for the meso-temperature
 
\begin{equation}
    \dot{T}_m = {a} \dot{\varepsilon}_v^2 + \frac{2}{3}{c} \dot{e}_{ij}\dot{e}_{ij} - \eta T_m^2,
    \label{eq:Tm_dot}
\end{equation}
 
where in order to retrieve the expression given in \citep[Equation 36][]{wiebicke2024simple}, we use 
 
\begin{equation}
    a^d = {a}, \quad \quad b^d_{ij} \equiv 0_{ij},\quad \quad     c^d_{ijkl} \equiv \frac{2}{3} {c} \delta_{ik}\delta_{jl},
\end{equation}
 
where $a$ and $c$ are positive material constants.

Using the above expressions and following Wiebicke and Einav's calibration procedure, we obtain the following plastic transport coefficients for the tensorially general case: 
 
\begin{equation}
    a^p = \sqrt{\frac{\eta}{a}}\frac{1}{p_c(\phi)}, \quad 
    b_{ij}^p  =-\frac{3}{2} \frac{a^p}{M^2}\frac{s^e_{ij}}{p^e},\quad 
    c_{ijkl}^p = \frac{3}{2}\left(\sqrt{\frac{\eta}{c}}\frac{1}{M\omega  p_c(\phi) }+\frac{a^p}{M^2}\right)\delta_{ik}\delta_{jl}.
\label{eq:transport_coefficients_ap}
\end{equation}

Given the specific form of the transport coefficients, the full tensorial expressions for the viscous stress and the plastic strain rate become
 
\begin{subequations}
\begin{align}
\label{eq:viscous_stress_final}
\sigma^d_{ij} &= \frac{2T^m}{\Gamma}\left( {a} \dot{\varepsilon}_v\delta_{ij} +\frac{2}{3}{c} \dot{e}_{ij}\right),\\
    \dot{\varepsilon}^p_{ij}  &= \frac{T^m}{M^2 p_1 \phi^{\lambda} }\sqrt{\frac{\eta}{a}}\left[\left( M^2- \left(\frac{q^{e}}{p^e}\right)^2\right)\frac{p^e}{3} \delta_{ij} +\frac{3}{2}\sqrt{\frac{{a}}{{c}}}\frac{M}{\omega} s^e_{ij}\right].
\label{eq:dot_eps_pl_final}
\end{align}
\end{subequations}

It is worth noticing that when evaluated in terms of the volumetric and deviatoric invariants -- namely, $p^d, q^d$ and $\dot{\varepsilon}^p_v, \dot{\varepsilon}^p_s$ -- the above expressions yield the exact same form of the transport coefficients as those derived in the original triaxial formulation of Terracotta model. Moreover, note that the requirement for the positive semidefiniteness of the plastic transport coefficients coincides with that identified in \citep[Equation (53)][]{wiebicke2024simple}, namely
\begin{equation}
    \frac{q_e}{p_e}\leq M_o, \qquad M_o = M\sqrt{\sqrt{\frac{a}{c}}\frac{M}{\omega}+1}.
    \label{eq:Mo}
\end{equation}

\subsection{Specification of the Terracotta friction law}
\label{subsec:friction-law}
From the newly expressed tensorial form of the Terracotta model, we specify the boundary conditions of an isolated fault gouge and derive the corresponding friction law under direct shearing configuration. In so doing, we use plane strain conditions and adopt the following minimal Voigt notation:
 
\begin{equation*}
\renewcommand\arraystretch{1.1}
    \sigma_{ij} \rightarrow \begin{bmatrix} \sigma_n \\ \sigma_{\ell}\\ \tau \end{bmatrix}, \qquad \varepsilon^e_{ij} \rightarrow   \begin{bmatrix} \varepsilon^e_n \\ \varepsilon^e_{\ell}\\ \gamma^e \end{bmatrix},
\end{equation*}
 
with superscripts $n, \ell$ referring to the normal and longitudinal directions, respectively. The equivalence with the tensorial form is given by $\sigma_n=\sigma_{11}$, $\sigma_{\ell} = \sigma_{22} =\sigma_{33}$, $\tau = \sigma_{12}$ and $\varepsilon^e_n=\varepsilon^e_{11}$, $\varepsilon^e_{\ell} = \varepsilon^e_{22} =\varepsilon^e_{33}$, $\gamma^e = 2\varepsilon^e_{12}$. 

Next, we consider a fault gouge of constant length $L$ and variable thickness $h$, whose state is determined by the following initial conditions at thermodynamic equilibrium
 
\begin{equation}
\renewcommand\arraystretch{1.1}
    h (0) = h^0, \quad \phi(0) = \phi^0, \quad T_m (0) = 0, \quad \varepsilon^e_{ij}(0) = \begin{bmatrix}
        \varepsilon^{e^0}_n\\ \varepsilon^{e^0}_{\ell}\\ 0\end{bmatrix}, \quad 
        \sigma_{ij}(0) = \begin{bmatrix}
        \sigma^{e^0}_n\\ \sigma^{e^0}_{\ell}\\ 0\end{bmatrix},
        \label{eq:initial_state}
\end{equation}
 
where the initial conditions for the elastic strain are obtained by equating the elastic stress in Equation (\ref{eq:elastic_stress}) with the prescribed values for $\sigma^e_n, \sigma_{\ell}^e$, under either isotropic or uniaxial stationary compression.

Starting from the initial state in Equation (\ref{eq:initial_state}), the fault gouge is subjected to a slip velocity $v$, see Figure \ref{fig:friction}(d). To ensure a constant normal stress $\sigma_n = \sigma_n^0$, the material undergoes dilation or contraction along the normal direction, following the evolution of the gouge thickness $h$ and its compaction velocity, $v_n$, again see Figure \ref{fig:friction}(d). With that, the resulting Terracotta friction law reads 
 
\begin{equation}
 \tau \Big(\Overbrace{\text{\footnotesize state}}{\varepsilon^e_v, \gamma^e, T_m, h}, \Overbrace{\text{\footnotesize rate}}{v^{\,^{\,}}, v_n} \Big) = \frac{A}{h^{6}} \varepsilon_v^e \gamma^e + \frac{B}{h}T_m v
    \label{eq:tau_specified}
\end{equation}
 
where the constants $A\equiv \tilde{G}\left( \phi^0h^0\right)^6$ and $B\equiv (2/3) c/\Gamma$. The above equation highlights the main result of the paper, which is formulated in terms of the following evolution equations for the state variables
 
\begin{subequations}
\label{eq:evolution_eqs}
\begin{align}
\dot{\varepsilon}_v^e&= \dot{\varepsilon}_n -\frac{T_m p^e }{ p_c(\phi) } \sqrt{\frac{\eta}{a}}\left(1-\left(\frac{q^e}{Mp^e} \right)^2\right) ,\\
\dot{\gamma}^e & =\dot{\gamma} - \frac{3T_m  \tau^e}{M\omega p_c(\phi)}\sqrt{\frac{\eta}{c}},\\
\dot{T}_m &= \left(a+\frac{4c}{9}\right) \dot{\varepsilon}_n^2+\frac{c}{3} \dot{\gamma}^2-\eta T_m^2,   \\
\dot{h} &= -h\dot{\varepsilon}_n,
\end{align}
\end{subequations}
 
where $\phi = (\phi^0 h^0)/h$. The total strain rates are then given in terms of the normal and slip velocities that act as the rate variables in the present model:
 
\begin{subequations}
\label{eq:rates}
\begin{align}
\dot{\gamma} &= \frac{v}{h} \; \text{ such that } v= v(t),\\
\dot{\varepsilon}_n &= -\frac{v_n}{h} \; \text{ such that } \dot{\sigma}_n \left(
    \text{state}, v, v_n \right) = 0,
    \label{eq:vn}
\end{align}
\end{subequations}
 
where $v(t)$ is imposed as function of time. \ref{appendix} presents further details of the numerical implementation of the friction law \citep[cf.][]{github}.

Furthermore, \ref{appendix_ss} presents an analytic derivation of the steady-state friction stress $\tau_s$ and coefficient $\mu_s$, under constant slip velocity $v$ and normal stress $\sigma_n$. Unlike conventional rate‐and‐state friction laws that consider a purely viscous model with logarithmic dependence on slip velocity, the Terracotta friction law embodies an inherent elastic component, proportional to the critical state line, $\propto M$, and a viscous term, proportional to the square of the velocity, $\propto v^2/h_s$, where $h_s$ is the gouge thickness at steady-state. Although the quadratic nature of the steady-state friction law limits the applicability of the current simple Terracotta friction law over extremely broad velocity ranges, as opposed to empirical logarithmic-like scaling in laboratory experiments, the new law provides robust predictive capabilities within the velocity regimes considered in the present analyses. In the applications to earthquakes of the Terracotta-spring model, the steady state develops under relatively small velocities, rendering the steady state limit of the friction law to be quite realistic. Nevertheless, it is envisaged that the Terracotta friction law could be reconciled with a logarithmic-like scaling by altering the current linear scaling of the rheological coefficients (i.e., $p_c$, $a$, $b$ and $c$) on the meso-temperature $T_m$.

In the following, the shear stress evolution predicted by the Terracotta friction law is expressed in terms of the friction coefficient, i.e., $\mu = \tau/\sigma_n$.
\subsubsection{Calibration of material parameters}
The Terracotta friction law (Equations \ref{eq:tau_specified}-\ref{eq:rates}) requires the specification of nine material parameters ($\tilde{K}$, $\tilde{G}$, $M$, $\omega$, $\lambda$, $\phi_I$, $a$, $c$, and $\eta$). However, when compared to conventional rate-and-state friction laws, which in their simplest form involve four to five parameters, the calibration of these parameters follows relatively simple and standard tests \citep{wiebicke2024simple} and account for much broader physics. For example, the two elastic constants, $\tilde{K}$ and $\tilde{G}$, allow us to describe pressure and density dependent elastic responses and can be calibrated using conventional elastic wave measurements \citep[cf.][]{riley2023consistent} or by fitting the model to observed stress-strain responses, as performed hereafter. Parameters $\phi_I$ and $\lambda$, which define inelastic compressibility that also goes beyond conventional both rate-and-state and micromechanically-inspired friction laws, can be directly calibrated from isotropic/uniaxial compression tests or inferred from the post-peak slope in triaxial compression or direct shear tests. The two critical state parameters, $M$ and $\omega$, can be obtained from triaxial compression or direct shear tests, with $\omega=0.5$ being a good value in the absence of density measurements, in agreement with the tacit assumption for clay in the modified cam clay model \citep{wood1990soil} of soil mechanics. The remaining three rheological parameters, $\eta$, $a$, and $c$, require calibration thorough rate-dependent experiments such as velocity stepping or slide-hold-slide tests, as those considered below.

\subsubsection{Calibration of system parameters}
Together with the above material parameters, it is necessary to determine appropriate initial conditions for the normal stress, $\sigma_n^0$, solid fraction, $\phi^0$, and layer thickness, $h^0$. The normal stress is conventionally prescribed or measured in laboratory experiments \citep[cf.][]{saffer2003comparison}, or in the field may be estimated as the overburden stress from the fault gouge depth \citep[cf.][]{kanamori2004physics}.
The solid fraction in general varies through the thickness of the fault gouge \citep[see][]{https://doi.org/10.1029/2022JB025666}, though as discussed earlier all such gradients are neglected in the present analysis.
One way to determine the initial layer thickness is to follow \citet{rice2006}, who associated it with the localised shear zone within the gouge of less than $1\div5$ mm. This shear zone of an ultra-cataclastic gouge is influenced by factors such as fault maturity and the chemical composition of its materials. It is worth noticing that in the present formulation, we consider a nominal uniform shear strain rate across the gouge thickness, which neglects the coseismic evolution of shear localisation from a full gouge thickness to a localised band. Nevetheless, despite precluding this important feature \citep[cf.][]{rice2006,https://doi.org/10.1002/2013JB010711,barras2025shear}, this simplification readily enables to account for the dependence of the frictional response of the shear zone thickness, cf. Sections \ref{sec:validation} and \ref{sec:spring-slider}, in a straightforward way. Further accounting for the dynamic evolution of the shear zone profile by solving the underlying boundary value problem can, in future, provide deeper insights into the transition from distributed to localised slip.

In clay, on which the current paper focuses, the thickness of the localised shear zones in the fault gouge is considered to be approximately $10\div 100$ times the clay platelet diameter, which is of the order of $1$ \textmu m  \citep{morgenstern1967microscopic,rice2006,haines2013shear,bigaroni2023frictional,volpe2024frictional}. Although studies have explored the effects of earthquake nucleation and seismic events on the layer thickness using both laboratory samples and natural fault gouges \citep[e.g.,][among others]{byerlee1976note,93JB03361,2011JB008264,evans1990thickness,JB095iB05p07007,scuderi2014physicochemical,lyu2019mechanics,bedford2021role}, the quantification of this relationship remains elusive. Therefore, the impact of layer thickness on the nucleation and seismicity is carefully analysed in Section \ref{sec:spring-slider}.

\section{Terracotta friction law: performance and validation}
\label{sec:validation}
The purpose of the current section is to demonstrate the capabilities of the proposed Terracotta friction law in capturing the complex frictional behaviour of clay-rich fault gouges. To this end, the analysis will make use of a variety of experiments on clay gouges conducted by \citet{saffer2003comparison}, \citet{bedford2022fault}, and \citet{ashman2023effect}, which explore the shear response of gouge media under different normal stress values and controlled velocity steps, as well as the evolution of the gouge's thickness $h$. 

As a start, we determine the corresponding material and system parameters of the Terracotta friction law. In particular, an initial guess is made based on previous values obtained by \citet{wiebicke2024simple}, which are then iteratively updated to fit the experimental curves, as specified below for each series of tests. The initial conditions are set according to the available experimental values of the normal stress, $\sigma_n^0$, and gouge's thickness, $h^0$. In the absence of information regarding the thickness of potential shear band, the gouge thickness, $h^0$ is assumed to match the initial thickness of the tested samples. While selecting a smaller value for $h^0$ would result in higher shear strain rates and alter the rheological parameters, the conceptual conclusions should remain valid for the scenarios considered here. 
Furthermore, since in the relevant papers information was not available concerning the initial values of the solid fraction $\phi^0$, the following analysis will consider it as a fitting parameter against the experimental results. Finally, the elastic strain state is initialised by isotropically compressing the simulated sample from a reference stress-free configuration, see Equation (\ref{eq:initial_state}). Table \ref{table:1} presents the model parameters and initial conditions for the gouge materials considered hereinafter. 
\begin{table}[ht]
\centering
  \begin{adjustbox}{max width=\textwidth}
\begin{tabular}{l c c c c c c cc c c } 
& \multicolumn{6}{c}{Mechanical parameters} & & \multicolumn{3}{c}{Rheological parameters}\\
\cline{2-7} \cline{9-11}
\vspace{-10pt}\\
& $\tilde{K}$ & $\tilde{G}$ & $M$ & $\omega$ & $\lambda$& $\phi_I$ & &  $a$ & $c$ & $\eta$ \\
 Gouge & (MPa) & (MPa) & (-)& (-)& (-)& (-) & & (Ks) & (Ks) & (K\textsuperscript{-1}s\textsuperscript{-1})\\
 \hline
 Illite & 1040 & 400 & 0.96 & 0.6 & 10& 0.20 && 7 & 7 &7.2$\cdot$10\textsuperscript{4} \\
Smectite & 260 & 100 & 0.42 & 0.5 & 10& 0.28 &&  2.4$\cdot$10\textsuperscript{4} & 2.4$\cdot$10\textsuperscript{4} &10\\
 Kaolinite & 15.6 & 12.5 & 0.44 & 0.4 & 4.5 & 0.05 && 0.42& 10&  3$\cdot$10\textsuperscript{6}\\
\hline
\end{tabular}
\end{adjustbox}
\caption{Parameters for the Terracotta friction law calibrated for the experiments conducted by Saffer \& Marone (2003) and Bedford et al. (2022) for illite shale, smectite, and kaolinite.}
\label{table:1}
\end{table}

\subsection{Experiments from Saffer and Marone (2003)}
\label{subsec:Marone}
The tests conducted by \citet{saffer2003comparison} consist of shearing of synthetic and natural gouges, in a double-direct shear geometry, at room humidity and temperature. Specimens were subjected to different levels of constant normal stress and sheared with varying velocities ranging from 1 to 200 $\mu$m/s. Focusing on clay-rich gouges, here we only study the tests on illite shale and smectite. The gouge layers were prepared by applying an initial compaction to the original sample resulting in an initial thickness $h^0$ equal to $2.25$ and $2.75$ mm for smectite and illite shale, respectively. Next, the gouge was sheared with frequent velocity stepping.  

The initial conditions for the Terracotta friction law consider the effective normal stress prescribed in the experiments, i.e. $\sigma_n = 20$ MPa, the aforementioned values for $h^0$, while the solid fractions were chosen as $\phi^0 = 0.57, 0.71$ for the illite and smectite gouges, respectively.

\begin{figure}[ht]
\centering
\includegraphics[width=0.56\textwidth]{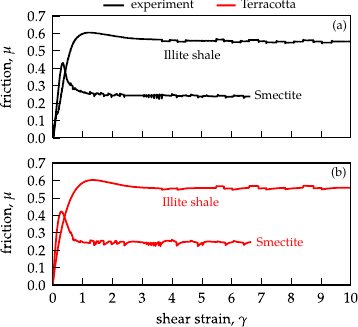}
\caption{Shear response of illite shale and smectite gouges under constant normal stress $\sigma_n = 20$ MPa, shown in terms of the friction coefficient, $\mu=\tau/\sigma_{n}$, versus the shear strain, $\gamma$: (a) experiments (Saffer \& Marone, 2003), and (b) model's predictions.}
\label{fig:Marone1}
\end{figure}

Figure \ref{fig:Marone1} displays the frictional behaviour during the experiments and the Terracotta friction law given the parameters in Table \ref{table:1} for the corresponding smectite and illite shale tests. Both the experiments and the predictions are characterised by an initially rapid increase in the friction coefficient during the loading phase, before peak friction coefficient at around $\gamma\approx 0.5\div 1$. Post-peak the friction coefficient gradually decays towards a plateau, whose value depends on the gouge's material, exemplified by the critical state constant, $M$, and the shear velocity.

The Terracotta friction law shows a strong capacity in replicating the responses of both specimens, including not only the  under velocity stepping during mature shearing, but also the peak and pre-peak values of the friction coefficient. The initial elastic response is considered to be entirely due to the elasticity of the gouge material since the stiffness of the loading apparatus is much larger \citep[cf. Section 3.3 in][]{saffer2003comparison}. Despite the very rich and variegate velocity dependence of the friction coefficient observed in the experiments, the new friction law is able to predict and distinguish between the large and limited velocity strengthening behaviour displayed by illite and smectite, respectively. The ability of the new law to adequately represent both pre- and post-peak behaviours moves beyond previously proposed rate-and-state and micromechanically-inspired frictional laws, where the elasticity of the gouge is ignored.

Some discrepancies remain between the experiments and the simulations in the amplitude of the friction coefficient jumps. While the experiments display a predominantly symmetrical trend of the friction coefficient with respect to up- and down-steps of the slip velocity, $v$, the predictions show jumps of $\mu$ more pronounced along one direction (up-steps for illite, and down-steps for smectite). The reason lies in the evolution of the meso-temperature, with a quadratic decay back to its steady-state value, cf. Equation (\ref{eq:Tm_dot}). This yields a corresponding quadratic dependence of the friction coefficient on the slip velocity, see \ref{appendix_ss}, making it difficult to accurately model the friction evolution under velocity steps of multiple orders of magnitude. Nevertheless, the predictions are found to be in good agreement with the experiments for the entire range of variation of the shear strain, in spite of the complex loading scenarios and the idealised boundary conditions (cf. direct shear).  

\begin{figure}[ht]
\centering
\includegraphics[width=\textwidth]{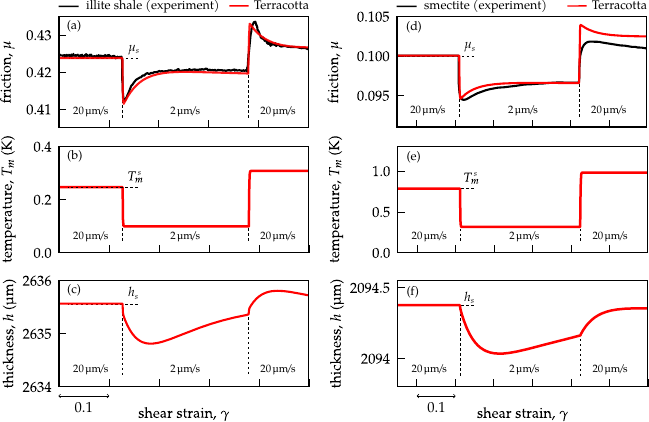}
\caption{Detailed view of velocity steps on illite shale (a-c) and smectite (d-f) under constant normal stress $\sigma_n=100$ MPa: (a,d) friction coefficient, $\mu$ (experiments after Saffer \& Marone, 2003) and model's predictions, (b,e) gouge's thickness, $h$, and (c,f) meso-temperature, $T_m$ predicted by Terracotta friction law. The subscript and superscript $s$ in $\mu_s$, $h_s$, and $T_m^s$ denote values at steady-state.}
\label{fig:velocity_step}
\end{figure}

To further illustrate the performance of the model under velocity-stepping conditions, Figure \ref{fig:velocity_step} shows a detailed comparison of the Terracotta predictions for experimental results on illite shale and smectite. Here, the gouges were sheared under constant normal stress $\sigma_n=100$ MPa, and subjected to a tenfold increase of the velocity till a critical state has been achieved, followed by a tenfold decrease of the velocity  \citep[cf. Figures 6 and 7 in][]{saffer2003comparison}. Calibration of the material parameters for this particular protocol yields the following values: $c=45\cdot 10^6$ Ks, $a\equiv c$, ${\eta} = 53\cdot 10^3$ K\textsuperscript{-1}s\textsuperscript{-1}, and $M=0.86$ for illite, and $c=11\cdot 10^7$ Ks, $a\equiv c$, ${\eta} = 13\cdot 10^3$ K\textsuperscript{-1}s\textsuperscript{-1}, and $M=0.21$ for smectite, while all remaining parameters are kept as those listed in Table \ref{table:1}. 

The Terracotta friction law successfully describes the velocity-strengthening behaviour of the gouges, with only some minor differences in the predictions for smectite. Figure \ref{fig:velocity_step} also presents the evolution of the meso-temperature $T_m$ and the gouge thickness $h$. The meso-temperature responds almost instantaneously to changes in slip velocity, shifting toward a new stationary value and causing a peak in the friction coefficient before it relaxes to a new steady-state. A similar pattern is observed in the gouge thickness, which undergoes compaction under velocity down-steps and dilation under velocity up-steps.

\subsection{Experiments from Bedford et al. (2022)}
\citet{bedford2022fault} performed experiments on fault gouge materials within a triaxial deformation apparatus in a direct-shear arrangement, involving both fully homogeneous and laterally heterogeneous gouges of kaolinite, quartz, and mixtures of them. The gouges, with an initial thickness $h^0\approx 1$ mm, were subjected under a constant (effective) normal stress $\sigma_n = 40$ MPa and sheared under relatively slow velocities between $0.3\leq v\leq 3$ \textmu m/s. 

\begin{figure}[ht]
\centering
\includegraphics[width=0.56\textwidth]{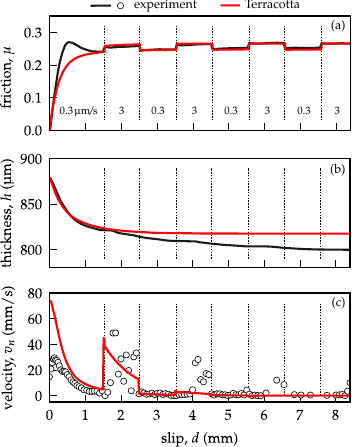}
\caption{Comparison between experiments (Bedford et al. 2022) and the Terracotta friction law for kaolinite at normal stress $\sigma_n=40$ MPa. Evolution of (a) friction coefficient, $\mu$, (b) gouge thickness, $h$, and (c) normal velocity, $v_n$.}
\label{fig:Lapusta}
\end{figure}

Here, we only focus on the clay-rich experiment that involved kaolinite, as presented in Figure \ref{fig:Lapusta}. The initial thickness and normal stress are set to match the experimental values, while the initial solid fraction is chosen as $\phi^0 = 0.65$.  This corresponds to an overconsolidation ratio of $p_c/p=1$, where $p=\sigma_n$ at the initial state.

As for the previous tests, the evolution of the friction coefficient during the slip was provided, as well as the evolution of the gouge thickness. The latter was tracked by measuring the volume of fluid expelled from the sample by means of servo-control pumps, which revealed a substantial amount of consolidation. In particular, the inference of the thickness was made while assuming constant sliding area, such that the volumetric straining was approximated through changes only to the layer thickness without orthogonal straining. Importantly, the data of the thickness variations allows us to corroborate the volumetric nature of the Terracotta friction law.

The new friction law is calibrated based on the time evolution of the gouge thickness, $h$, presented in Figure \ref{fig:Lapusta}(b), before the application of velocity steps that initiates after slip displacement of $1.5$ mm.
The comparison of the model against the experimentally measured friction coefficient, gouge thickness, and normal velocity, $v_n$, is presented in Figure \ref{fig:Lapusta}. Despite an underestimation of the peak value of the friction coefficient, the predictions beyond this peak are found to be in good agreement with the experimental results, and demonstrate capacity to capture the isotach phenomenon \citep{suklje1969rheological} at both up- and down-steps of the velocity. The evolution of the gouge thickness is well captured as well, thanks to the adopted calibration procedure. In particular, the model agrees with the experiments until the first downstep of the velocity, around $2$ mm slip, where the friction law predicts a near constant sample thickness, in contrast with the slight continual compaction observed in the experiments. It is worth noticing that the intrinsic volumetric nature of the new friction law depends on the time evolution of two rates -- slip velocity $v(t)$ as well as compaction velocity $v_n(t)$. To this purpose, we also inferred the experimental evolution of $v_n$, by means of a first-order finite difference scheme from the test data of the thickness, i.e., $v_n(t) \approx v(t)\Delta h/\Delta d$, where $\Delta h$ and $\Delta d$ are the increments of the gouge's thickness and slip displacement. As shown in the correspond19ing Figure \ref{fig:Lapusta}(c), the predicted normal rate, $v_n$, well matches the experiments. Despite some minor differences mostly due to the resolution of the recorded evolution of the thickness, the Terracotta friction law correctly predicts the values of the compaction velocity before and during velocity stepping, with minor underestimation of the repeated upsteps of the shear velocity.

Finally, it is worth noticing that the minor differences between the predictions and the experiment may be due to the working assumption of material homogeneity in the current initial development. A further reason may be due to the idealised boundary conditions considered in the specification of the new friction law (see subsection \ref{subsec:friction-law}) as compared to the actual setup of the present experiment \citep{bedford2022fault}, where  soft silicone spacers were positioned at end of the gouge layers in order to accommodate longitudinal displacements without supporting any load. 

\subsection{Experiments from Ashman and Faulkner (2023)}
The experiments performed by \citet{ashman2023effect} use the same experimental setup of \citet{bedford2022fault}, again on kaolinite gouges in order to study their frictional behaviour under velocity steps. However, in their study they additionally inspected the influence of the normal stress, $\sigma_n$. Accordingly, we consider the exact same material and rheological parameters obtained from the calibration of the previous test, as listed in Table \ref{table:1}, and vary the initial solid fraction to keep a constant overconsolidation ratio of $p_c/p=1$.

\begin{figure}[H]
\centering
\includegraphics[width=0.56\textwidth]{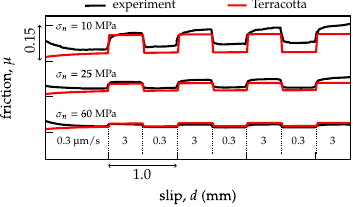}
\caption{The variations in the friction coefficient, $\mu$, of kaolinite in response to seven slip velocity steps under different constant normal stresses of $\sigma_n = 10, 25, 60$ MPa, in terms of experiments (Ashman \& Faulkner, 2023), and the Terracotta friction law. The friction coefficient curves are plotted at the same scale, around a mean value of $\mu =0.33$, with an offset of 0.15 along the y-axis for clarity.}
\label{fig:faulker}
\end{figure}

Figure \ref{fig:faulker} compares the predictions provided by the Terracotta friction law against the experimental results. An overall good agreement is found, which is notable considering that no adjustment of the material parameters was made for the different normal stresses thanks to the dependence of the viscous component of the steady-state friction coefficient on $\sigma_n$ (cf. \ref{appendix_ss}). In particular, the decreasing amplitude of the isotachs at increasing normal stress is accurately predicted by the new friction law, with only minor differences visible under the least stressed configuration, i.e., $\sigma_n=10$ MPa.

\subsection{Parametric analysis}
\label{subsec:parametric}
This paragraph presents a parametric study that demonstrates the capacity of the Terracotta friction law in recovering a rich spectrum of rate‐and‐state responses. The setup mirrors the velocity‐step protocol of Figure \ref{fig:velocity_step}(a) for illite, under identical initial and boundary conditions. Here, the parametric study systematically explores the roles of (i) the rheological coefficients and (ii) the elastic moduli, while focusing on their influence on the steady‐state response and transient behaviour during velocity jumps.

\subsubsection{Impact of the rheological coefficients}
\label{subsubsec:rheo_param}
\begin{figure}[ht]
\centering
\includegraphics[width=1\textwidth]{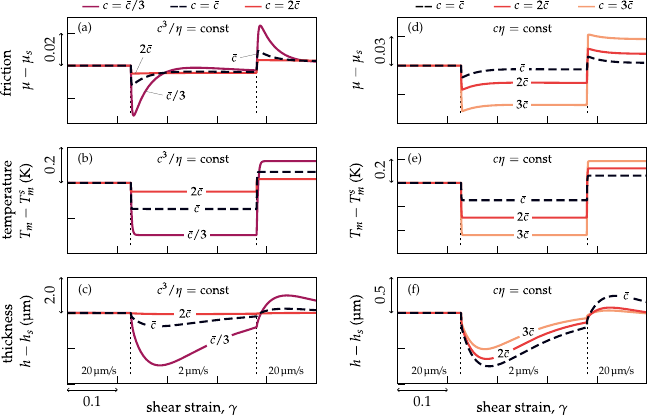}
\caption{The impact of the rheological parameters on the model response under velocity steps (cf. Figure 5). To this end, $c=a$ and $\eta$ are varied according to two protocols. The first considers $c^3/\eta = \bar{c}^3/\bar{\eta}$ with $\bar{c}=45\cdot 10^6$ Ks, $\bar{\eta} = 53\cdot 10^3$ K\textsuperscript{-1}s\textsuperscript{-1}): (a) friction, (b) meso-temperature, and (c) gouge's thickness. The second maintains $c\eta = \bar{c}\bar{\eta}$ constant: (d) friction, (e) meso-temperature, and (f) gouge's thickness. All variables are shown as offsets with respect to the steady-state values preceding the first velocity jump for each simulation (refer to Figure 5).}
\label{fig:velocity_step_parametric}
\end{figure}

To understand the phenomenological significance of the rheological constants, $c$ (or $a$) and $\eta$, we perform two sets of tests, while holding all other parameters fixed. In the first set, Figure \ref{fig:velocity_step_parametric}(a-c), we use a combination of constants $c$ ($=a$) and $\eta$ that maintains the same stationary viscous stress at a given strain rate by keeping $c\propto \eta^{1/3}$. In the second set, Figure \ref{fig:velocity_step_parametric}(d-f), we vary $c$ and $\eta$ while maintaining $c\propto \eta^{-1}$ which affects the stationary viscous stress.

It is shown that by enforcing the proportionality $c\propto \eta^{1/3}$, the stationary viscous component of the friction coefficient is maintained constant during the initial loading phase, allowing any subsequent variations in the relaxation rate to be attributed exclusively to differences in the transient evolution of the meso-temperature. 
 Figure \ref{fig:velocity_step_parametric}(a-c) demonstrates that higher values of $c$ and $\eta$ accelerate the relaxation of the transients of the meso-temperature and friction coefficient, while increasing the magnitude of compaction/dilation of the gouge layer, without perturbing the early-time response. 

Conversely, when $c$ and $\eta$ are varied while keeping the product $c\eta$ fixed, the magnitude of the friction coefficient jumps increases with increasing $c$ due to a greater component of the viscous stress, as shown in Figure \ref{fig:velocity_step_parametric}(d-f). This in turn affects the underlying compaction and dilation of the gouge. The shape of the transient response to velocity jumps is not affected as the transient of the meso-temperature is kept identical during the loading. 

These analyses demonstrate the ability of the Terracotta friction law to predict diverse rate-dependent phenomena \citep[see also][]{wiebicke2024simple}. The model predicts the characteristic behaviour displayed by illite with rapid transitions of the stress between constant plateaus under step changes in the strain rate, a phenomenon often referred to as `isotach' behaviour \citep{suklje1969rheological} in soil mechanics. The model also predicts the instantaneous jumps of the shear stress followed by a relaxation towards a new residual rate-dependent plateau as shown by smectite. Those rate-dependent phenomena are governed by one of the new state variables introduced in Terracotta, the meso-temperature. This is in contrast with classical rate-and-state frictional laws \citep{dieterich1979modeling,dieterich1981constitutive,ruina1983slip} where the form of the rate-dependence is described in terms of (fixed) parameters ($a$ and $b$). Here, the meso-temperature influences the range of predicted rate-dependent behaviours. 

\begin{figure}[ht]
\centering
\includegraphics[width=1\textwidth]{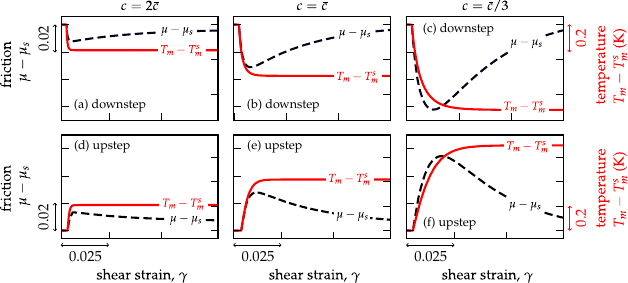}
\caption{The role of the meso-temperature in establishing the short-term, peak friction response. Here, the rheological parameters satisfy $c^3/\eta = \bar{c}^3/\bar{\eta}$, with different $c=a$ and $\eta$, as per Figure 8(a): (left column) $c=2\bar{c}$, (centre) $c=\bar{c}$, and (right) $c=\bar{c}/3$. Subfigures (a-c) refer to the velocity downstep and subfigures (d-f) display the velocity upstep.}
\label{fig:velocity_step_parametric2}
\end{figure}

To better examine how $T_m$ controls the transient frictional response, Figure \ref{fig:velocity_step_parametric2} presents magnified views of the evolution of $T_m$ and $\mu$, corresponding to the parametric study with $c\propto \eta^{1/3}$ that was shown in Figure \ref{fig:velocity_step_parametric}(a-c). Although relative to the friction coefficient, $T_m$ rapidly attains its steady-state value (see Figure \ref{fig:velocity_step_parametric}), the rate at which it evolves -- governed by its evolution equation (\ref{eq:Tm_dot}) -- directly controls the time taken to establish the peak friction coefficient and consequently its value. For decreasing values of $c$, the transient dynamics of $T_m$ prolongs. As a result, we observe longer durations for the friction coefficient to establish its peak and increased peak values. Eventually, well after $T_m$ reaches its stationary value, the friction coefficient slowly relaxes due to the progressive reduction of the viscous stress.

\subsubsection{Impact of the elastic coefficients}
\label{subsubsec:elastic_param}
\begin{figure}[ht]
\centering
\includegraphics[width=\textwidth]{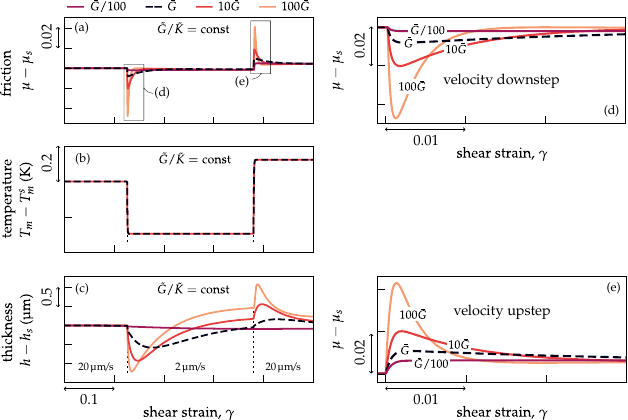}
\caption{The impact of the elastic parameters on the model response under velocity steps (cf. Figure 5). To this end, $\tilde{G}$ and $\tilde{K}$ are varied by ensuring $\tilde{G}/\tilde{K} = \bar{G}/\bar{K}$, with $\bar{G}=400$ MPa and $\bar{K} = 1040$ MPa: (a) friction, (b) meso-temperature, and (c) gouge's thickness. Subfigures (d) and (e) present magnified plots of the friction coefficient during the velocity lower and higher jumps, respectively. All variables are shown as offsets with respect to their respective steady-state values preceding the first velocity jump (a-e) and the second velocity jump (f-g).}
\label{fig:velocity_step_KG}
\end{figure}

Next, we explore how the elasticity of the gouge through its elastic stiffness moduli, $\tilde G$ and $\tilde K$, influence the response. By maintaining the ratio $\tilde{G}\propto \tilde{K}$ fixed, we keep the critical stress ratios between $M_{e}$, $M_0$ and $M$ unchanged \citep{wiebicke2024simple}, which ensures the requirements expressed in Equations (\ref{eq:Me}) and (\ref{eq:Mo}) are met in all the simulations. Figure \ref{fig:velocity_step_KG}(a–c) displays the friction coefficient, the meso‐temperature, and the gouge's thickness for both velocity‐down and velocity‐up steps. Panels (d–g) show zooms on the transient phases.

Increasing $\tilde G$ and $\tilde K$ results in larger magnitudes of the friction jumps and faster relaxations toward the new steady state. The physical elasticity of the material thus provides a new meaning to the extent of the so-called `critical slip distances' that are commonly lumped into a phenomenological constant in classical rate-and-state friction models (cf. $D_c$). By enforcing the proportionality $\tilde{G}\propto \tilde{K}$, the steady-state value and magnitude of the jumps of the meso-temperature remain unchanged. Instead, the gouge thickness evolution is strongly affected by the elasticity of the system: with higher jumps under increased elastic moduli $\tilde{G}$ and $\tilde{K}$. According to the present model, the extent of dilation (compaction) is a direct signature of the elasticity of the system, further suggesting a potential indirect measure of gouge elasticity in experiments. 

\section{Terracotta-spring model for earthquakes dynamics}
\label{sec:spring-slider}
This section extends the analysis to investigate the predictive capabilities of the Terracotta friction law in modelling earthquake nucleation, stick-slip instabilities, and their occurrence. We focus attention on exploring the influence of the physical parameters that are unique to this new friction law, such as slip thickness, meso-temperature, and solid fraction, and which cannot be effectively examined using conventional rate-and-state models \citep{kanamori2004physics,dieterich1979modeling,dieterich1981constitutive,ruina1983slip,pipping}. To this end, we connect the Terracotta friction law to a spring-dashpot representation of the host rock, as shown in Figure \ref{fig:spring-slider}(c). We then use this combined Terracotta-spring model to analyse the statistical dynamics of earthquakes and its energetical terms.
In this model the mass, $m$, represents a mobilised mass sliding over a frictional gouge layer of thickness $h$ and length $L$, undergoing a slip displacement $d$ due to the far-field movement of the tectonic plates, which is prescribed by the slip velocity $v_{\infty}$. The mass is connected to a spring with stiffness $k={G}_* L$, which represents the effective mobilised stiffness of the host rock, where ${G}_*$ is the corresponding shear modulus. This spring is connected in parallel to a dashpot with damping coefficient $\varsigma$, equivalent to the apparent viscosity of the host rock. Subjected to an effective overburden normal stress $\sigma_n$, the frictional shear layer develops a shear stress $\tau$, governed by the evolution of Terracotta's state variables and the slip velocity, $v=\dot{d}$, see subsection \ref{subsec:friction-law}.

\subsection{Governing equations}
Using force balance, the motion of the Terracotta-spring model is described as follows
 
\begin{equation}
\label{eq:spring-slider}
    m\dot{v} = k(d_{\infty}-d)+\varsigma ( v_{\infty} - v) - S\tau,
\end{equation}
 
where $m=\rho_* L^3$ is the mobilised mass assuming a cuboid of length $L$, $S=L^2$ is the contact area, $d_{\infty} = v_{\infty}t$ is the displacement at the extremity of the spring-dashpot configuration, and $\varsigma = 2\zeta \omega_n m$, where $\zeta$ is the damping ratio and $\omega_n = \sqrt{\sfrac{{G}_*}{\rho_*}}/L$ is the natural frequency of the system, assuming an equivalent linearly elastic isotropic medium for the host rock, while $\rho_*$ is its bulk density.

Next, we introduce the following dimensionless variables
 
\begin{subequations}
\begin{align}
    \hat{t}&= \frac{t}{\mathscr{T}}, \quad \gamma=\frac{d}{D}, \quad \gamma_{\infty} = \frac{d_{\infty}}{D}, \quad \varepsilon_n = 1-\frac{h}{D},\\
    \gamma' &= \frac{v}{v_{\infty}}, \quad \varepsilon_n' = - \frac{v_n}{v_{\infty}}, \quad \theta = \frac{T_m}{\Theta},
\end{align}
\end{subequations}
 
where $D$, $\mathscr{T}$, and $\Theta$ are respectively the characteristic length, time, and meso-temperature, i.e.,
 
\begin{equation}
    D = h^0, \quad \mathscr{T} = \frac{D}{v_{\infty}}, \quad \Theta =  \frac{1}{\mathscr{T}} \sqrt{\frac{c}{3\eta}}.
\end{equation}

Here, $D$ coincides with the initial value of the fault gouge thickness, $\mathscr{T}$ denotes the timescale over which shear strain accumulates within the gouge, and $\Theta$ corresponds to the ambient meso-temperature -- that is, the steady-state value reached under a constant slip velocity $\mathscr{T}^{-1}$ and constant dilation/contraction of the fault gouge. Accordingly, the non-dimensional forms of Equations (\ref{eq:spring-slider}) and (\ref{eq:tau_specified}) read
 
\begin{subequations}
\label{eq:dimensionless}
\begin{align}
    \label{eq:dimensionless_spring_slider}
    \gamma'' &= \hat{k} \left(\gamma_{\infty} - \gamma\right) +\hat{\varsigma} \left(1 -\gamma'\right) - \hat{N} \mu, \vspace{10pt}\\
    \mu &= \frac{\hat{A}}{\left(1-\varepsilon_n\right)^6}\varepsilon_v^e\gamma^e+\frac{ \hat{B}}{1-\varepsilon_n}\theta\gamma',
    \label{eq:dimensionless_friction}
\end{align}
\end{subequations}
 
with
 
\begin{equation}
\begin{split}
    \hat{k} &= \frac{\epsilon^2}{\mathscr{C}}, \quad \hat{\varsigma} = 2\zeta \sqrt{\hat{k}} , \quad \hat{N} = \frac{\hat{k}}{\epsilon \kappa}, \\
    \hat{A} &= {\phi^{\vspace{-4pt} 0}}^6\frac{\tilde{G}}{\sigma_n}, \quad \hat{B} =  \mathscr{T}^{-1}\Theta\frac{2c}{3\Gamma\sigma_n},\\
        \epsilon &= \frac{D}{L}, \quad \mathscr{C} = \frac{\rho_* v_{\infty}^2}{{G}_*}, \quad \kappa = \frac{G_*}{\sigma_n},
\end{split}
\end{equation}
 
where $\mathscr{C}$ is the Cauchy number representing the ratio of inertial and elastic forces through the host rock and $\epsilon$ is the thickness-to-length ratio of the fault gouge.

The dynamics of the Terracotta-spring model is governed by six non-dimensional variables -- $\gamma$, $\varepsilon_n$, $\gamma'$, $\varepsilon_n'$, $\varepsilon^e_v$, $\gamma^e,$ and $ \theta$ -- in terms of the non-dimensional time $\hat{t}$. The model's behaviour is determined through the time integration of Equation (\ref{eq:dimensionless}), along with the evolution equations for the state variables $\varepsilon^e_v$, $\gamma^e$, $\theta$, and $\delta_n$, as presented in Equation (\ref{eq:evolution_eqs}). We recall that the evolution of the normal displacement, $\delta_n$, is prescribed by ensuring a constant stress $\sigma_n$. The initial conditions for these state variables are given in Equation (\ref{eq:initial_state}).

\section{Terracotta-spring model: results}
The previous section has described the attachment of the Terracotta friction law with a spring-dashpot system representative of the host rock. Here we analyse the corresponding performance for modelling earthquake dynamics. Besides the material constants associated with the friction law in subsection \ref{subsec:friction-law} and the initial conditions ($\sigma_n$, $h^0$, and $\phi^0$), five additional non-dimensional parameters of the spring-slider system must be specified: $\mathscr{T}$, $\epsilon$, $\mathscr{C}$, $\kappa$, and $\zeta$. Their values depend on the actual configuration of the fault gouge, either in laboratory experiments or in the field. To this end, in the following a parametric analysis is performed, as motivated by previous such explorations of rate-and-state friction laws \citep{rice1985constitutive,rudnicki1988physical,JB095iB05p07007,riceruina83,gu1991}.

Here, we consider a clay-rich gouge with material parameters similar to those calibrated using direct shear tests (see Section \ref{sec:validation}) and by triaxial experiments by \citet{wiebicke2024simple}. In particular, we select values in the range obtained from these calibrations \citep[see Section \ref{sec:validation} and][]{wiebicke2024simple}, namely: $0.016\leq K\leq 26$ GPa, $0.013\leq G\leq 13$ GPa, $0.42\leq M \leq 1.3$, $0.4 \leq \omega \leq 0.65$, $3.62\leq \lambda\leq 17.2$, $0.05\leq \phi_I \leq 0.45 $, $7 \leq c\leq 3\cdot 10^7$ Ks, and $0.01\leq\eta\leq 3\cdot 10^6$ K\textsuperscript{-1}s\textsuperscript{-1}.
The material and in-situ dimensional properties are given in Table \ref{table:2}, where the volumetric and deviatoric rheological constants are assumed equal, i.e., $a\equiv c$. The following analysis is presented in terms of a unit fault gouge segment, $L=1$ m, and constant stiffness to focus attention on the new and key ingredients of the Terracotta-spring model. In most scenarios, the results are being presented in their non-dimensional form to further enable scaling of the relationships identified in this work. Stability analyses of the present model are expected to gather additional insights in the occurrence and magnitude of the predicted stick-slip instabilities \citep[cf.][]{riceruina83,gu1984slip, gu1991,ciardo2024non}, but such investigation remains beyond the scope of the present work.

\begin{table}[ht]
\centering
\begin{adjustbox}{max width=\textwidth}
\begin{tabular}{@{\hskip 5pt}c@{\hskip 5pt} c@{\hskip 5pt} c@{\hskip 5pt} c@{\hskip 5pt} c@{\hskip 5pt} c@{\hskip 5pt} @{\hskip 5pt}c@{\hskip 5pt} c@{\hskip 10pt} c@{\hskip 5pt} c@{\hskip 5pt} c@{\hskip 5pt} c@{\hskip 5pt}@{\hskip 5pt}c@{\hskip 5pt} c@{\hskip 5pt}  } 
\multicolumn{8}{c}{Material parameters} & & \multicolumn{4}{c}{In-situ properties}\\
\cline{1-8} \cline{10-13}
\vspace{-8pt}\\
$\tilde{K}$ & $\tilde{G}$ & $M$ & $\omega$ & $\lambda$ & $\phi_I$ & $c$ & $\eta$ & $\quad$ & $G_*$  & $\rho_*$ & $\zeta$ & $v_{\infty}$\\
(GPa) & (GPa) & (-) & (-) & (-) & (-) & (Ks) & (K\textsuperscript{-1}s\textsuperscript{-1}) & & (GPa) &  (kg/m\textsuperscript{3}) & (-) & (\textmu m/s)\vspace{3pt}\\
\cline{1-8} \cline{10-13}
\vspace{-8pt}\\
18.2 & 7 & 0.9 & 0.6 & 10 & 0.225 & $10^5$ & $0.01$ & & 10  & 2500 & $0\div 1$ & $10^{-5}\div 50$\vspace{3pt}\\
\cline{1-8} \cline{10-13}
\end{tabular}
\end{adjustbox}\\
\medskip

\begin{adjustbox}{max width=\textwidth}
\begin{tabular}{@{\hskip 5pt}c@{\hskip 5pt} c@{\hskip 5pt} c@{\hskip 5pt} } 
\multicolumn{3}{c}{Initial conditions} \\
\cline{1-3}
\vspace{-8pt}\\
$\sigma_n$ & $\phi^0$ & $h^0$ \\
(MPa) & (-) & (mm) \\
\cline{1-3}
\vspace{-8pt}\\
$\; 5\div 640\; $ & $\; 0.52 \div 0.9\; $ & $\; 0.001\div 10\; $ \\
\cline{1-3}
\end{tabular}
\end{adjustbox}
\caption{Baseline parameters for the Terracotta-spring model per metre unit length of a fault gouge segment. The rheological parameters are assumed to be equal, i.e., $a\equiv c$. The far-field tectonic velocity, $v_{\infty}$, varies between $1$ mm/year ($\sim 3\cdot 10^{-5}$ \textmu m/s) and $50$ \textmu m/s.}
\label{table:2}
\end{table}

\subsection{Benchmark}
\label{subsec:benchmark}
\begin{figure}[ht]
\centering
\includegraphics[width=\textwidth]{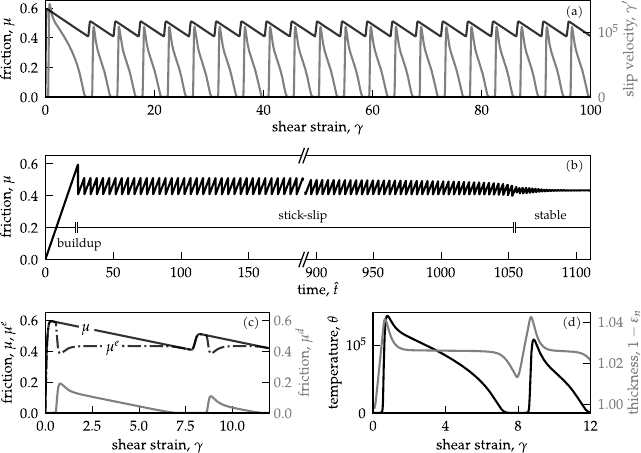}
\caption{Response of the Terracotta-spring model: (a) friction coefficient $\mu$ and non-dimensional slip velocity (or non-dimensional strain rate) $\gamma'$ as functions of the shear strain $\gamma$; (b) friction coefficient $\mu$ during the non-dimensional time $\hat{t}$ (b); (c) a focus view on the evolution of various components of the friction coefficient; and (d) a focus view highlighting the synchronicity of the meso-temperature $\theta$ and the non-dimensional thickness, $1+\varepsilon_n$. Results refer to $v_{\infty} = 1$ cm/year, and initial conditions: $h^0 = 0.1$ mm, $\phi^0 = 0.71$, and $\sigma_n = 40$ MPa.}
\label{fig:result1}
\end{figure}

To demonstrate the predictive capabilities of the Terracotta-spring model, we consider a benchmark scenario where the parameters are set as follows: $v_{\infty} = 1$ cm/year (equivalent to $\sim 3\cdot 10^{-4}$ \textmu m/s), $\zeta=0$ (undamped scenario), and the following initial conditions: $h^0 = 0.1$ mm, $\phi^0 = 0.71$, and $\sigma_n = 40$ MPa. The initial solid fraction corresponds to an overconsolidation ratio equal to $p_c/p\sim 2.6$. Figure \ref{fig:result1} presents the corresponding response of the Terracotta-spring analogue model, including the evolution of the friction coefficient, $\mu$, and non-dimensional slip velocity (or non-dimensional strain rate), $\gamma'$, as functions of the slip deformation, $\gamma$, and the non-dimensional time, $\hat{t}$, as well as the evolution of the meso-temperature, normal strain, slip, and friction coefficient.

Figures \ref{fig:result1}(a,b) showcase three key behaviours: (1) the initial elastic buildup of the friction coefficient, (2) the nucleation of earthquakes triggered by frictional instability with sustained stick-slip behaviour, and (3) a transition to smooth and steady sliding. This sequence begins with an isolated seismic event followed by persistent and periodic stick-slip limit cycles. The first event is characterised by a stress drop of $7.4$ MPa (with a friction drop $\Delta \mu \sim 0.18$) accompanied by a fault slip of $0.7$ mm occurring in $\sim 30$ s, reaching a maximum slip velocity $v>10^5 v_{\infty}$. The subsequent periodic events display stress drops of about $4$ MPa, with $0.4$ mm slip displacements developing over $\sim 20$ s. These periodic events have a return period of $\sim 16$ days. The stick-slip behaviour gradually converges to a steady, aseismic slip around $\hat{t} = 1000$, equivalent to $10$ years, with the chosen non-dimensionalisation.

Examining Figures \ref{fig:result1}(b-d), the proposed friction law is effectively predicting the nucleation of earthquakes, during which the slip velocity increases until the inertia starts dominating and cause the Terracotta layer to accelerate to a peak velocity after which the friction drops dynamically. The friction drop can be decomposed into two contributions, elastic and viscous, which follows the relation $\tau=\tau^e+\tau^d$ based on the derivation in Section \ref{sec:model}. The elastic component of the friction, $\mu^e = \tau^e/\sigma_n$, initially drops sharply, while the gouge compacts (decreasing $\varepsilon_n$), which leads to higher slip velocities and causes the viscous component, $\mu^d = \tau^d/\sigma_n$, to increase. The elastic component then rapidly rebuilds under constant normal strain, reaching a stationary value as the viscous component diminishes. At this juncture, the system appears to shift into an emergent stability, where the material compacts leading to an additional decrease of the friction. Throughout the stick-slip cycles, the model predicts persistent dilation-compaction cycles of the fault gouge, associated with increase-decrease of the shear zone thickness, $1+\varepsilon_n$ (see Figure \ref{fig:result1}(d)). Additionally, the evolution of the meso-temperature, $\theta$, follows a similar pattern to the slip velocity, but also accounts for the fluctuating motion at the mesoscale related to the normal strain rate, $\varepsilon_n'$, and the energy flowing towards the microscale.

In summary, the proposed Terracotta-spring model enables to study the nucleation of earthquakes, with the development of well-defined and persistent stick-slip sliding, which is synchronised with the evolution of the meso-temperature and the evolution of the shear zone thickness, with an intermittent pattern. To investigate the transition from stick-slip cycle to smooth sliding, cf. Figure \ref{fig:result1}(b), we considered a critically-damped scenario ($\zeta=1$). In this case, the model yields the exact same dynamic behaviour, where persistent stick-slip cycles develop, with only a time-increasing lag in the recurrence of the periodic seismic events of the order of $0.03$ at $\hat{t}=100$. Therefore, in the following we shall consider the model without the dashpot using $\zeta=0$, for simplicity.

\subsection{Influence of physical parameters}
Here, we investigate the influence of physical parameters and namely the slip zone thickness, the solid fraction, the normal stress, and the far-field velocity. While the influence of some of these parameters has been extensively studied using previous rate-and-state and micromechanically-inspired friction laws, as well as using high-fidelity simulations including finite element and discrete element methods, the effects of other physical parameters remain challenging to capture with simplified one-dimensional models. The objective here is twofold: (1) to examine the impact of key physical quantities which were not well studied in the past, such as slip zone thickness, and (2) to assess whether the newly proposed friction law exhibits trends consistent with those observed in experiments but also in conventional rate-and-state friction laws, thereby validating its relevance in capturing known frictional mechanisms.

\subsubsection{Slip zone thickness, $h$}
In contrast with conventional state-of-the-art friction models that treat friction as a surface phenomenon, a key feature of the Terracotta friction law is the incorporation of a finite slip zone thickness. In the Terracotta-spring model, the thickness evolves during both the earthquake nucleation and slip, and its initial value appears to significantly influence the frictional instabilities, limit cycles, and stress drops associated with each seismic event. It is also worth noticing that the current model does not account for phenomena such as particle breakage, thermal pressurization, flash heating, or frictional melting, each of which is thought to influence the shear zone's thickness and fault surface roughness evolution \citep[for an extensive review, see][]{faulkner2010review,collins2020cosserat,stathas2023fault}. Nonetheless, the results shed light on additional important phenomena that complement those listed above.

\begin{figure}[h!]
\centering
\includegraphics[width=\textwidth]{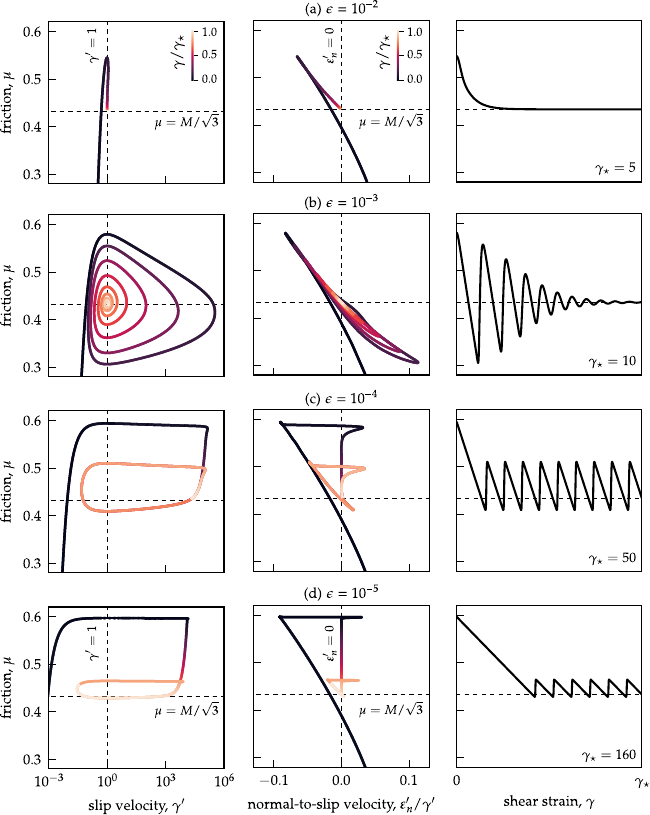}
\caption{Influence of the non-dimensional gouged shear zone thickness, $\epsilon$, on earthquake dynamics. Evolution of the friction coefficient with respect to the non-dimensional slip velocity (left), the normal-to-slip velocity (centre), and shear strain (right) under decreasing shear zone thickness (from top to bottom): (a) $\epsilon=10^{-2}$, (b) $\epsilon=10^{-3}$, (c) $\epsilon=10^{-4}$, and (d) $\epsilon=10^{-5}$. Results refer to slip velocity of $v_{\infty} = 1$ cm/year and initial conditions of $\phi^0 = 0.71$ and $\sigma_n = 40$ MPa.}
\label{fig:phase_h}
\end{figure}

Following the experimental evidence collected and presented by \citet{rice2006}, we select initial thicknesses between $10 \text{ \textmu m} \geq h^0\geq 10$ mm. The parametric study is conducted per metre unit length of the fault gouge segment, leading to a non-dimensional thickness parameter $10^{-5} \geq \epsilon \geq 10^{-2}$. Figure \ref{fig:phase_h} shows the phase portraits (left and centre) of the friction coefficient with respect to the slip velocity, $\gamma'$, and the ratio of normal-to-slip velocity, $\varepsilon_n'/\gamma'$, as well as the friction coefficient as a function of the accumulated slip (right) for decreasing thicknesses (from top to bottom). For the relatively largest thickness, i.e., $\epsilon=10^{-2}$ ($h^0= 10$ mm), the model predicts an isolated seismic event followed by a smooth, steady sliding. When the initial thickness decreases to $\epsilon=10^{-3}$ ($h^0=1$ mm), the model exhibits stick-slip behaviour with rapidly decreasing amplitudes, which ultimately converge to the same fixed-point attractor as in the $\epsilon=10^{-2}$ case. The attractor coincides with the steady-state friction (see \ref{appendix_ss}), which is approximately given by $\mu \sim M/\sqrt{3}$ owing to the relatively low slip velocities developed. For even smaller initial thicknesses, i.e., $\epsilon\leq 10^{-4}$ ($h^0\leq 100$ \textmu m), the model predicts sustained stick-slip limit cycles with amplitudes inversely related to the thickness. In this latter scenario, the limit cycles also appear in the normal-to-slip velocity, demonstrating that the coupling between frictional slip and the underlying compaction/dilation of the fault gouge actively participates in the dynamics of earthquakes.

\begin{figure}[h]
\centering
\includegraphics[width=\textwidth]{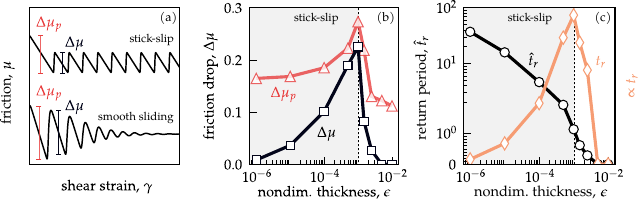}
\caption{Influence of the non-dimensional gouged shear zone thickness, $\epsilon$, on earthquake characteristics. From left to right: (a) identification of the friction drops, $\Delta \mu_p$ and $\Delta \mu$, for stick-slip and smooth sliding, (b) evolution of the friction drops, and (c) return period, $\hat{t}_r$ ($t_r$), as s function of $\epsilon$. Results refer to slip velocity of $v_{\infty} = 1$ cm/year and initial conditions of $\phi^0 = 0.71$ and $\sigma_n = 40$ MPa.}
\label{fig:dmu_h}
\end{figure}

The results are further analysed and summarised in Figure \ref{fig:dmu_h} which shows the evolution of the friction drops and their recurrence intervals as function of the gouged shear zone thickness. As this thickness decreases, the non-dimensional recurrence period of the seismic events $\hat{t}_r$ increases. However the physical (dimensional) recurrence interval, $t_r$, decreases at the decreasing of the thickness within the stick-slip region. The friction drops are separated into two categories: the primary, typically larger drop, $\Delta \mu_p$, and subsequent drops, $\Delta \mu$ that, in the case of stick-slip behaviour, correspond to the limit cycle amplitude (also refer to Figure \ref{fig:result1}). The results suggest that there is a critical shear zone thickness that produces the largest stress drop. For larger $\epsilon$, the friction drops decrease rapidly as stick-slip transitions to steady, aseismic sliding. Similarly, as $\epsilon$ decreases, the friction drops for subsequent limit cycles approach zero, reflecting the decreasing amplitude of stick-slip events, while the primary drop reaches an asymptote, $\Delta \mu_p\sim 0.16$, independently of $\epsilon$.

It is worth noticing that predicted transition between steady slip and stick-and-slip behaviours is a direct result of the coupled evolution of the meso-temperature and the volumetric nature of the Terracotta friction law, under the imposed direct-shear boundary conditions. Different initial gouge's thicknesses affect the evolution of the meso-temperature and its steady-state value (cf. \ref{appendix_ss}) which, in turn, alters the evolution of the thickness and thus the solid fraction.

The above results gather new insights on how stick-slip behaviour and nucleation of earthquakes are strongly related to the shear zone thickness. Note that the negative proportionality in the friction drops and return intervals at increasing initial thicknesses is in agreement with laboratory test of synthetic granular fault gouges \citep{lyu2019mechanics}.

\subsubsection{Solid fraction, $\phi$}
\begin{figure}[h]
\centering
\includegraphics[width=\textwidth]{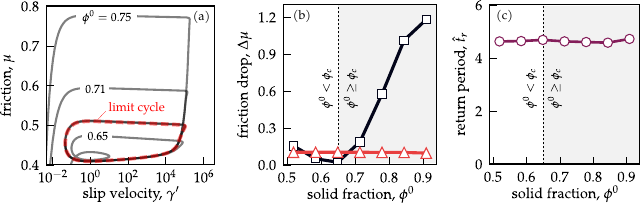}
\caption{Influence of the solid fraction, $\phi^0$, on earthquake dynamics and characteristics under constant normal stress $\sigma_n=40$ MPa: (a) phase portrait of the friction versus the non-dimensional slip velocity and limit cycle, (b) friction drops, and (c) return period in terms of $\phi^0$. Results refer to $v_{\infty} = 1$ cm/year and initial conditions of $h^0 = 0.1$ mm shear band thickness. The value of the solid fraction corresponding to the normally consolidated scenario is $\phi_c = 0.65$.}
\label{fig:dmu_phi}
\end{figure}

Next, we examine the impact of the initial solid fraction on the seismic response. Figure \ref{fig:dmu_phi} presents these results in terms of the $\mu-\gamma'$ limit cycles, the magnitude of friction drops, and the return interval.  Figure \ref{fig:dmu_phi}(a) shows that the limit cycle attractor remains independent of $\phi^0$, leading to a constant $\Delta \mu$ and return interval $\hat{t}_r$, respectively in Figures \ref{fig:dmu_phi}(b,c). However, the initial solid fraction influences the primary friction drop, with two different trends emerging. For $\phi^0$ higher than the threshold for a normally consolidated gouge, $\phi_c$, the initial friction drop $\Delta \mu_p$ increases with increasing $\phi^0$, where $\phi_c\sim 0.65$ is obtained by requiring $p_c(\phi)/p=1$.
This aligns with the concept of critical state in soil mechanics, where denser materials display a more pronounced stress peak followed by strain softening before reaching critical state. In contrast, for looser states ($\phi^0\leq \phi_c$), the friction drop increases as $\phi^0$ decreases, due to stress adjustments during the elastic buildup that precedes the primary drop and enabling the compaction of the gouge.

\subsubsection{Normal stress, $\sigma_n$}
\begin{figure}[h]
\centering
\includegraphics[width=\textwidth]{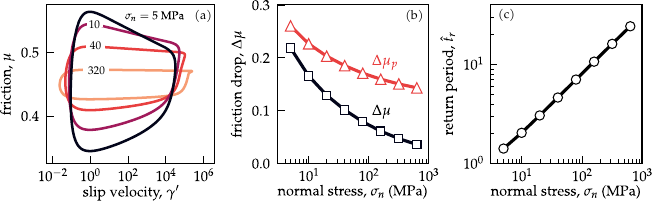}
\caption{Influence of the normal stress, $\sigma_n$, on earthquake dynamics and characteristics under a constant overconsolidation ratio $p/p_c\sim 4$: (a) stick-slip limit cycles, (b) friction drops, and (c) return period in terms of $\sigma_n$. Results refer to $v_{\infty} = 1$ cm/year, and initial conditions of $h^0 = 0.1$ mm.}
\label{fig:dmu_sn}
\end{figure}

The results in Figure \ref{fig:dmu_sn} shows responses for various constant normal stress, $\sigma_n$ from 5 to 640 MPa. As $\sigma_n$ decreases, both the primary and cyclic friction drops increase, in agreement with experimental evidence \citep{saffer2003comparison,numelin2007frictional}. In particular, Figure \ref{fig:dmu_sn}(a) shows that the limit cycles tend to expand toward higher cyclic friction values and contract at lower slip velocities as $\sigma_n$ decreases. The return period instead increases exponentially with increasing normal stress.

\begin{figure}[h]
\centering
\includegraphics[width=0.7\textwidth]{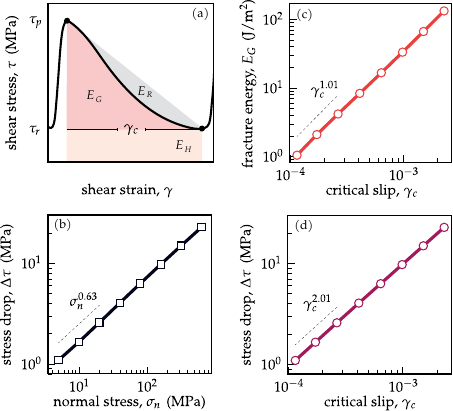}
\caption{Influence of the normal stress on fault energetics and statistics: (a) schematic representation of the energy budget; (b) stress drop as a function of the normal stress; and (c,d) fracture energy and stress drop over $\gamma_c$ defined in (a).}
\label{fig:dmu_sn_wow}
\end{figure}

To further assess the predictive capabilities of the Terracotta-spring model relative to other models, simulations, and data from natural and induced seismicity, we temporarily move from the non-dimensional analysis to consider physical variations of dimensional parameters. Figure \ref{fig:dmu_sn_wow}(a) shows an example of a slip-weakening curve obtained from the new model and definition of energetic parts of the budget per unit area during a seismic event, for $\sigma_n = 40$ MPa. As conventionally referred to in geophysics, the energy budget comprises the three contributions explained in Figure \ref{fig:dmu_sn_wow}(a) and identified as: (1) released elastic energy, $E_R$, (2) fracture energy, $E_G$, and (3) frictional energy, $E_F$ \citep{kanamori2004physics,kanamori2006energy}. For $\sigma_n=40$ MPa, we find $E_G = 0.21$ J/m\textsuperscript{2}, $E_F=170$ J/m\textsuperscript{2}, and $E_R = 0.04$ J/m\textsuperscript{2}.

Figure \ref{fig:dmu_sn_wow}(b) shows a power law increase of the cyclic stress drops  with normal stress, which well compares with previous discrete element method numerical simulations and experimental studies on gouges \citep{passelegue2016dynamic,bedford2021role,bai2022numerical}. Here, we get $\Delta \tau \propto \sigma_n^{0.63}$.
Furthermore, the model predictions for the relationships between the stress drops, $\Delta \tau$, and fracture energy, $E_G$, with respect to the critical slip, $\gamma_c$, align well with prior experimental studies conducted on smooth and gouge faults \citep{aubry2018frictional,passelegue2016dynamic,ohnaka2003constitutive,doi:https://doi.org/10.1002/9781119156895.ch12,https://doi.org/10.1029/2001JB000670}, rock fracturing \citep{yoshimitsu2014magnitude}, acoustic emission data \citep{goodfellow2014laboratory}, and discrete element method numerical simulations \citep{kato2012dependence,bai2022numerical,https://doi.org/10.1029/2022JB025666}. Both $\Delta \tau$ and $E_G$ well match a power law with slip: $\Delta \tau\propto \gamma_c^{1.016}$ and $E_G\propto \gamma_c^{2.008}$. A first-order approximation of the fracture energy, given by the triangular area $(\tau_p - \tau_r)/\gamma_c$, suggests $E_G \propto \Delta \tau \gamma_c = \gamma_c^{2.016}$, which indicates minimal released energy, $E_R$.

\subsubsection{Far-field velocity, $v_{\infty}$}
\begin{figure}[h]
\centering
\includegraphics[width=\textwidth]{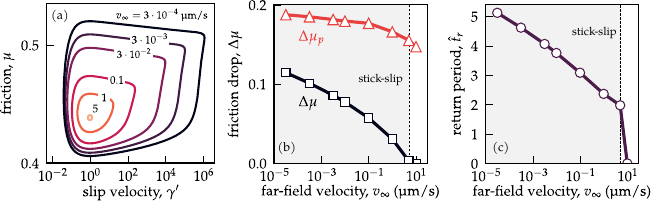}
\caption{Influence of the far-field velocity, $v_{\infty}$ on fault energetics and statistics: (a) stick-slip limit cycles, (b) friction drops, and (c) return period versus $v_{\infty}$. Results refer to initial conditions of $h^0 = 0.1$ mm, $\phi^0 = 0.71$, and $\sigma_n = 40$ MPa.}
\label{fig:dmu_vinf}
\end{figure}

Finally, we study the influence of the tectonic far-field velocity, $v_{\infty}$, by testing over six order of magnitude variations, i.e., $10^{-5}\leq v_{\infty} \leq 50$ \textmu m/s. Figure \ref{fig:dmu_vinf} summarises the results which are in agreement with the well-known dependence of fault gouges on $v_{\infty}$ \citep{scholz2019mechanics,kanamori2004physics,gu1991,gu1994nonlinear}.

The far-field velocity determines the loading rate of the fault, and is found to directly influence the return period and indirectly affect the stress drops and stick-slip limit cycles. Higher velocities lead to faster stress accumulation, producing more frequent seismic events of lower magnitude. On the contrary, lower velocities yield slower loading rates, which in turns produce larger return periods, with seismic events characterised by larger stress drops. This is true both for the primary stress drop, $\Delta \mu_p$, and the persistent periodic drops, $\Delta \mu$. It is worth noticing that at vary large velocities, the stick-slip limit cycles tend to converge to a fixed-point attractor at $\mu \sim M\sqrt{3}$.

\begin{figure}[h]
\centering
\includegraphics[width=0.34\textwidth]{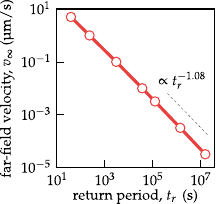}
\caption{Influence of the far-field velocity, $v_{\infty}$: relationships between $v_{\infty}$ and return period, $t_r$.}
\label{fig:dmu_vinf2}
\end{figure}

The return interval and the far-field velocity are found to follow a power law, i.e., $v_{\infty} \propto t_r^{-1.08}$, as shown in Figure \ref{fig:dmu_vinf2} and in agreement with existing natural seismicity data and laboratory tests \citep{beeler2001earthquake}.

\section{Conclusions}
This study develops a new pathway to study the frictional behaviour of fault gouges that is grounded on hydrodynamic principles from theoretical physics~\citep{landau2013statistical} that considers the way energy flows in heterogeneous media through the use of thermal and meso-related temperatures \citep{jiang2009granular,einav2023hydrodynamics}. In particular, we built upon the hydrodynamic Terracotta model of clay \citep{wiebicke2024simple}. Initially formulated for triaxial loading, this paper established its tensorially generalised form, in order to solve the particular loading conditions of directly sheared fault gouges. Unlike previous models, this pathway ensures an objective evaluation that does not engineer the friction law against observational trends. As a result, the new friction law accounts for the volumetric response of the materials, and directly captures the dependence of fault gouge dynamics on thickness, normal stress and solid fraction, as well as explains for the first time analytically the buildups of an elastic stress towards first rupture.

The performance of the Terracotta friction law was validated against laboratory experiments of clay-rich fault gouges across various scenarios, including velocity stepping, varying normal stress, and gouge compaction \citep{saffer2003comparison, ashman2023effect, bedford2022fault}. By integrating this friction law with a conventional spring-dashpot representation of the host rock,  the full seismic cycle was simulated, showcasing both shear stress accumulation and earthquake nucleation. The numerical results of this Terracotta-spring model successfully captured several well-established trends observed in laboratory experiments, advanced simulations, as well as natural and induced seismicity. Key result includes the model predictions of the finding by~\citet{lyu2019mechanics} of the dependence of friction drops and return intervals on the initial gouge thickness. Furthermore, the same Terracotta-spring model offers insights into behaviours that await experimental studies, including: (1) the existence of a critical shear zone thickness that maximises stress drops; and (2) the independence of the return periods on the initial solid fraction of the gouge.

The newly developed friction law represents a significant step beyond phenomenological rate-and-state laws, in that it enriches understanding of the physics of friction of fault gouges. Thanks to the physical generality of the hydrodynamic procedure \citep{landau2013statistical,jiang2009granular} and the simplicity of Terracotta \citep{wiebicke2024simple}, the use of the new friction law should be applicable to model general heterogeneous geomaterials, beyond just clay-rich gouges as done in this paper. As show in \citep{einav2023hydrodynamics,chen2023hydrodynamic}, the hydrodynamic procedure can accurately describe multi-physics coupled phenomena, which have important implications on fault dynamics \citep[see][among others]{rice2006,veveakis2012modeling,rattez2018importanceI,barbot2022rate,stathas2023fault}.

While in this paper the initial thickness of the fault gouge was taken as a parameter, the evolution equation of the meso-temperature in the complete Terracotta model showcases dependence on its second gradient that naturally introduces an evolving length scale to continuum mechanics. With that, it is anticipated that the Terracotta constitutive model could help addressing both the genesis of faults -- through the localisation of strain, kinetic fluctuations (meso-temperature), and thermal pressurisation \citep[among others, see][]{stathas2023fault} -- and the subsequent evolution of rate- and state-dependent earthquake cycles.

\appendix
\section{Numerical implementation}
\label{appendix}

The numerical implementation of Terracotta friction law, Equations (\ref{eq:tau_specified}-\ref{eq:rates}), is developed in Python, by solving the initial value problem specified by the evolution equations for the state variables and the time derivative of the shear stress, Equation (\ref{eq:tau_specified}), given the initial conditions in Equation (\ref{eq:initial_state}), using the \texttt{solve\_ivp} package from \texttt{SciPy} library \citep{2020SciPyNMeth}. The time derivative of the shear stress is obtained by time differentiation of Equation (\ref{eq:tau_specified}): 
\begin{equation}
    \dot{\tau} = \frac{\partial \tau}{\partial \varepsilon_v^e}\dot{\varepsilon}_v^e+\frac{\partial \tau}{\partial \gamma^e}\dot{\gamma}^e+\frac{\partial \tau}{\partial T_m}\dot{T}_m+\frac{\partial \tau}{\partial h}v_n+\frac{\partial \tau}{\partial v}\dot{v},
\end{equation}
where $\dot{\varepsilon}_v^e$, $\dot{\gamma}^e$, $\dot{T}_m$, and $v_n$ (or $\dot{h}$) are prescribed by Equations (\ref{eq:evolution_eqs}) and (\ref{eq:rates}). The slip acceleration, $\dot{v}$, is approximated with a backward finite-difference scheme. 

The solution of the initial value problem is driven by the prescribed time evolution of the shear strain rate, $\dot{\gamma}$ (or equivalently, the slip velocity, $v$) and the imposed value of the normal stress, $\sigma_n$.
To guarantee a constant value of the normal stress, a root-finding algorithm is used to identify the normal-strain rate, $\dot{\varepsilon}_n$ (or, $v_n$), to yield
\begin{equation}
    \dot{\sigma}_n = \frac{\partial {\sigma}_n}{\partial \varepsilon_v^e}\dot{\varepsilon}_v^e+\frac{\partial {\sigma}_n}{\partial \varepsilon_s^e}\dot{\varepsilon}_s^e+\frac{\partial {\sigma}_n}{\partial \varepsilon_n^e}\dot{\varepsilon}_n^e+\frac{\partial {\sigma}_n}{\partial T_m}\dot{T}_m+\frac{\partial {\sigma}_n}{\partial h}v_n+ \frac{\partial {\sigma}_n}{\partial v_n}\dot{v}_n=0,
    \label{eq:sigman_dot}
\end{equation}
where the rates of the state variables are functions of the state, as well as the shear and normal strain rates. Equation (\ref{eq:sigman_dot}) is numerically solved using the \texttt{SciPy}'s \texttt{minimize} package.

Further notice that in all simulations, the numerical implementation of Equation (\ref{eq:vn}) uses the initial thickness, $h^0$, rather than the current thickness, $h$, for the sake of simplicity. This assumption is confirmed by the results where the normal strain is found to be infinitesimal small, i.e., $\lesssim 5 \%$.

The system of ordinary differential equations defining the initial value problem is numerically integrated using a fifth-order implicit Runge-Kutta method from the Radau IIA family \citep{2020SciPyNMeth}. Adaptive time stepping is employed, with absolute error tolerances varying between 10\textsuperscript{-8} and 10\textsuperscript{-12}, to ensure that no significant differences are observed in the computed solutions.

The numerical implementation of the Terracotta-spring model, Equations (\ref{eq:dimensionless}), is analogous to that of the Terracotta friction law, with the only difference that the slip velocity, $\dot{\gamma}$, and acceleration $\ddot{\gamma}$, are defined by the solution of the balance equation of the Terracotta-spring-dashpot system, cf. Equation (\ref{eq:dimensionless_spring_slider}). For more details, we refer to the code \citep{github}.

\section{Steady-state friction law}
\label{appendix_ss}
The steady-state limit of the Terracotta friction law can be found by imposing stationarity for the state variables under constant slip velocity, $v\geq 0$, and effective normal stress, $\sigma_n$. The stationarity of those variables is given when they (\ref{eq:evolution_eqs}) vanish:
\begin{equation}
    \dot{\varepsilon}_v^e = 0, \quad \dot{\gamma}^e = 0, \quad \dot{T}_m=0, \quad \dot{h}=0.
\end{equation}
Under the above conditions the stationary values of the meso-temperature and gouge's thickness become
\begin{equation}
    T_m^s = \sqrt{\frac{c}{3\eta}}\frac{v}{h_s}, \qquad h_s = \phi^0h^0\left( \frac{\omega p_I}{\sigma_n-T_m^{s^2}/\Gamma}\right)^{1/\lambda},
\end{equation}
where we considered most accurately and for all practical purposes $\sigma_n^{e}=p^{e}$ at steady-state, or equivalently that $\sigma^{e}_{\ell}=\sigma_n^{e}$.

Using the above relations, the steady-state friction stress reads
\begin{equation}
\tau_s \left(T_m^s, h_s \right)= \frac{M}{\sqrt{3}}\left( \sigma_n- \frac{T_m^{s^2}}{\Gamma}\right)+ B\sqrt{\frac{c}{3\eta }}\left(\frac{v}{h_{s}}\right)^2.
\end{equation}

Furthermore, due to the second order dependence of $T_m^s$ on the strain rate in all considered scenarios, the thermodynamic pressure, $T_m^{s^2}/\Gamma$, is negligible compared to the elastic normal stress. Accordingly, we arrive at the following simple expression for the friction law at steady state:
\begin{equation}
    \tau_s (h_s)\approx \frac{M}{\sqrt{3}} \left( \sigma_n +\frac{B}{M}\sqrt{\frac{c}{\eta}}  \left(\frac{v}{h_{s}}\right)^2\right),
\end{equation}
or equivalently for the steady-state friction coefficient
\begin{equation}
    \mu_s (h_s)\approx \frac{M}{\sqrt{3}} \left( 1 +\frac{B}{M\sigma_n}\sqrt{\frac{c}{\eta}}  \left(\frac{v}{h_{s}}\right)^2\right),
\end{equation}
with $h_s\approx \phi^0h^0\left( \omega p_I/\sigma_n\right)^{1/\lambda}$.

The steady-state friction consists of an elastic contribution, proportional to the critical state line, $\propto M$, and a viscous term, proportional to the square of the velocity and inversely proportional to the applied normal stress, $\propto v^2/(h_s \sigma_n)$, where $h_s$ is the gouge thickness at steady-state and $\sigma_n$ is the normal stress. This quadratic dependence could be tuned in the future by altering the current linear dependence of the Onsager's rheological coefficients on $T_m$.

\section*{Open Research}
\noindent The experimental data sets used for the validation of the model were obtained directly from the corresponding works \citep{saffer2003comparison,bedford2022fault,ashman2023effect}. The software for reproducing the results presented in this work is available at \citep{github}.

\section*{Acknowledgements}
The authors would like to acknowledge the support of the Australia Research Council (ARC) under the Discovery Projects scheme
(Grant agreement ID DP220101164: ``Physics-informed hydrodynamic model for clay across scales''). Special thanks are also extended to Dr. Max Wiebicke for valuable discussions about the developed model.

\begin{notation}
\notation{} \hspace{-9pt} Einstein’s summation is implied for repeated indices $i,j=1,2,3$
\notation{$a, b, c$} Volumetric, off-diagonal, and deviatoric plastic transport coefficients
\notation{$d_{\infty}, d$} Far-field and gouge's slip displacements
\notation{$D$} Characteristic length
\notation{$e^e_{ij} = \varepsilon^e_{ij} - \smash{\frac{\varepsilon^e_v}{3}} \delta_{ij}$} Deviatoric elastic strain tensor
\notation{$f_i, f^m_i$} Thermal and meso-related entropy currents
\notation{$\tilde{G},\tilde{K}$} Intrinsic shear and bulk moduli
\notation{$G_*$} Host rock shear modulus
\notation{$h$} Gouged shear zone thickness
\notation{$m$} Mobilised mass
\notation{$M$} Critical stress ratio
\notation{$L$} Gouge's length
\notation{$p =\smash{\frac{\sigma_{ii}}{3}}$} Total pressure
\notation{$p_c(\phi)$} Isotropic compression line
\notation{$p_I$} Constant defining the isotropic compression line
\notation{$p^d$} Viscous pressure
\notation{$p^e$} Elastic pressure
\notation{$p^T$} Thermodynamic pressure
\notation{$q=\smash{\sqrt{\sfrac{3}{2}\; s_{ij}s_{ij}}}$} Deviatoric total stress (invariant)
\notation{$q^d$} Viscous deviatoric stress
\notation{$q^e$} Elastic deviatoric stress
\notation{$R, R_m$} Thermal and meso-related entropy productions
\notation{$s, s_m$} Thermal and meso-related entropies
\notation{$s_{ij} = \sigma_{ij} - p\delta_{ij}$} Deviatoric stress tensor
\notation{$t$} Time
\notation{$\mathscr{T}$} Characteristic time
\notation{$\hat{t}$} Non-dimensional time
\notation{$T, T_m$} Thermal temperature and meso-temperature
\notation{$u, u^e, u^m$} Total, elastic, and meso-related internal energy densities
\notation{$v_i$} Velocity vector
\notation{$v, v_n, v_{\infty}$} Slip, normal, and far-field velocities
\notation{$\mathbb{X}^d_{ijkl},\mathbb{X}^p_{ijkl}$} Viscous and plastic transport coefficients\\

\notation{$\Box^0$} Initial value of variable $\Box$
\notation{$\dot{\Box} = \frac{\partial x}{\partial t}$} Derivative of variable $\Box$ with respect to time $t$
\notation{$\Box' = \frac{\partial \Box}{\partial \hat{t}}$} Derivative of the variable $\Box$ with respect to the non-dimensional time $\hat{t}$
\notation{$\nabla_i \square = \frac{\partial \Box}{\partial x_i}$} Gradient of the scalar variable $\Box$
\notation{$\nabla_i \square_i = \frac{\partial \square_i}{\partial x_i}$} Divergence of the vectorial variable $\square_i$
\notation{$\nabla^2_i \Box = \frac{\partial^2 \Box}{\partial x_i^2}$} Laplacian of the scalar variable $\Box$\\

\notation{$\gamma^e$} Elastic shear strain
\notation{$\dot{\gamma}, \dot{\gamma}^e$} Total and elastic shear strain rates
\notation{$\Gamma$} Fixed coefficient that sets the unit of the meso-temperature
\notation{$\bar{\eta}$} Energy sink coefficient
\notation{$\eta$} Constant parameter specifying the energy sink
\notation{$\delta_{ij}$} Kronecker delta
\notation{$\varepsilon^e_{ij}$} Elastic strain tensor
\notation{$\varepsilon^e_v =\varepsilon^e_{ii}$} Volumetric stress invariant
\notation{$\varepsilon_s^e=\sqrt{\sfrac{2}{3}\; \smash{e^e_{ij}e^e_{ij}}}$} Deviatoric elastic strain invariant
\notation{$\dot{\varepsilon}^e_{ij}, \dot{\varepsilon}_{ij}, \dot{\varepsilon}_{ij}^p$} Elastic, total, and plastic strain-rate tensors
\notation{${\varepsilon}^e_n, {\varepsilon}^e_{\ell}$} Normal and longitudinal elastic strains
\notation{$\varepsilon_n$} Normal strain
\notation{$\dot{\varepsilon}_n, \dot{\varepsilon}_{\ell}$} Normal and longitudinal strain rates
\notation{$\mu$} Friction coefficient
\notation{$\phi$} Solid fraction
\notation{$\phi_I$} Solid fraction of the isotropic compression line, at $p=1$ kPa
\notation{$\lambda$} Slope of the isotropic compression line
\notation{$\rho$} Bulk density
\notation{$\rho_s$} Solid density
\notation{$\rho_*$} Host rock's bulk density
\notation{$\sigma_{ij}$} Total stress tensor
\notation{$\sigma^d_{ij}$} Viscous stress tensor
\notation{$\sigma^e_{ij}$} Elastic stress tensor
\notation{$\sigma_n,\sigma_{\ell}, \tau$} Normal, longitudinal, and shear stresses
\notation{$\Theta$} Characteristic meso-temperature
\notation{$\omega$} Position of the critical state line
\notation{$\zeta$} Damping ratio
\end{notation}

\bibliographystyle{abbrvnat}
\bibliography{bibliography} 

\end{document}